\documentclass[aps,prb,reprint,twocolumn,superscriptaddress,floatfix,citeautoscript,longbibliography]{revtex4-2}
\usepackage[bookmarks=true,colorlinks,linkcolor=OrangeRed,urlcolor=NavyBlue,citecolor=RoyalBlue]{hyperref}
\usepackage{epsfig,amsmath,amssymb,color,comment,physics}
\usepackage[makeroom]{cancel}
\usepackage[caption=false,font=normalsize]{subfig}
\usepackage{mathrsfs}
\usepackage[countmax]{subfloat}
\usepackage[normalem]{ulem}
\usepackage[english]{babel}
\usepackage{dsfont}
\usepackage{placeins}%\Floatbarrier
\usepackage[dvipsnames]{xcolor}
\usepackage{svg} %include svg files
\usepackage{graphicx}% Include figure files
\usepackage{appendix}

\usepackage{multirow} %tables
\usepackage{graphicx}  %tables

\usepackage{bbold} %mathbb(1)
\usepackage{dcolumn}% Align table columns on decimal point
\usepackage{bm}
\usepackage[capitalise]{cleveref}
\begin{document}
\title{Dipolar ordering transitions in many-body quantum optics:\\
Analytical diagrammatic approach to equilibrium quantum spins}

\author{Benedikt Schneider}
\affiliation{Department of Physics and Arnold Sommerfeld Center for Theoretical
Physics, Ludwig-Maximilians-Universit\"at M\"unchen, Theresienstr.~37,
80333 Munich, Germany}
\affiliation{Munich Center for Quantum Science and Technology (MCQST), 80799 Munich, Germany}

\author{Ruben Burkard}
\affiliation{Institut für Theoretische Physik, Universit\"at T\"ubingen, Auf der Morgenstelle 14, 72076 T\"ubingen, Germany}

\author{Beatriz Olmos}
\affiliation{Institut für Theoretische Physik, Universit\"at T\"ubingen, Auf der Morgenstelle 14, 72076 T\"ubingen, Germany}

\author{Igor Lesanovsky}
\affiliation{Institut für Theoretische Physik and Center for Integrated Quantum Science and Technology, Universit\"at T\"ubingen, Auf der Morgenstelle 14, 72076 T\"ubingen, Germany}
\affiliation{School of Physics and Astronomy and Centre for the Mathematics and Theoretical Physics of Quantum Non-Equilibrium Systems,
The University of Nottingham, Nottingham NG7 2RD, United Kingdom}

\author{Bj\"orn Sbierski}
\affiliation{Institut für Theoretische Physik, Universit\"at T\"ubingen, Auf der Morgenstelle 14, 72076 T\"ubingen, Germany}

\begin{abstract}
Quantum spin models with a large number of interaction partners per spin are frequently used to describe modern many-body quantum optical systems like arrays of Rydberg atoms, atom-cavity systems or trapped ion crystals. For theoretical analysis the mean-field (MF) ansatz is routinely applied. However, besides special cases of all-to-all or strong long range interactions, the MF ansatz provides only approximate results. Here we present a systematic correction to MF theory based on diagrammatic perturbation theory for quantum spin correlators in thermal equilibrium. Our analytic results are universally applicable for any lattice geometry and spin-length $S$. We provide pre-computed and easy-to-use building blocks for Ising, Heisenberg and transverse field Ising models in the symmetry-unbroken regime. We showcase the quality and simplicity of the method by computing magnetic phase boundaries and excitations gaps. We also treat the Dicke-Ising model of ground-state superradiance where we show that corrections to the MF phase boundary vanish. 
\end{abstract}

\date{\today}
\maketitle

\maketitle

%%%%%%%%%%%%%%%%%%%%%%%%%%%%%%%%%%%%%%%%%%%%%%%%%%%%%%%%%
%%%%%%%%%%%%%%%%%%%%%%%%%%%%%%%%%%%%%%%%%%%%%%%%%%%%%%%%%
\section{Introduction}

The past decade has witnessed tremendous progress at the intersection points of cold atomic physics, quantum optics and many-body physics. Atoms and ions can be confined in spatially structured arrangements \cite{endresAtom-by-atom2016-1,barredoAtom-by-atom2016-1,BarreiroOpen-system2011,ShankarSimulating2022} and brought into interaction using tailored potentials or cavity-mediated forces \cite{friskKockumUltrastrong2019,BaumannDicke2010,LandigQuantum2016}. This has opened a new window for the exploration of equilibrium and non-equilibrium phenomena, including phase transitions in arrays of trapped Rydberg atoms \citep{browaeysManyBodyPhysics2020} and Wigner crystals of trapped ions \citep{monroeProgrammableQuantum2021,guoSiteresolvedTwodimensional2024}, super- and subradiance in dense atomic gases \citep{guerinSubRadiance2016,RuiSubradiantmirror2020} or exotic time-crystal phases in atom-cavity systems \citep{kongkhambutContinuousTimeCrystal2022,CabotContinuous2024}.

\begin{figure}[t]
\begin{centering}
\includegraphics{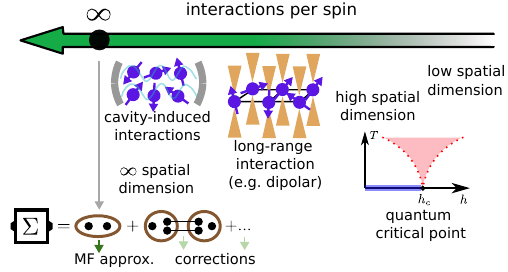}
\par\end{centering}
\centering{}\caption{\label{fig:JConnectivityOverview}The validity of the MF approximation
in spin systems depends on the number
of interactions per spin (green arrow) and is only exact in the limit
of infinite connectivity. Corrections to MF can be moderate in systems
if the connectivity of spins is large, a common situation in many-body
quantum optics, but also in high spatial dimensions or at quantum
critical points. In this regime the diagrammatic method proposed in
this work provides accurate results at negligible computational cost.}
\end{figure}

One commonality of these quantum optical platforms is that their essential physics is often captured by many-body models whose microscopic degrees of freedom are quantum spins.  
However, the structure of the underlying spin-spin couplings
is strikingly different to the short-range interactions encountered in solid-state based quantum magnetism \citep{auerbachInteractingElectrons1994}. 
%This is due to long-wavelength photons and phonons that mediate the interactions between atoms and ions, which thereby give rise to tunable long-range and even all-to-all spin-spin couplings 
This is due to the spatially extended nature of the mode-functions of photons (phonons) that mediate the interactions between atoms (ions) and thereby give rise to tunable long-range and even all-to-all spin-spin couplings
\citep{LarsonMott2008,kirtonIntroductionDicke2019,mivehvarCavityQED2021,SvendsenModifieddipole2023,leKienNanofiber2017}, see Fig.~\ref{fig:JConnectivityOverview}.

In this theoretical work we are concerned with the treatment of such highly-connected spin Hamiltonians, motivated but not limited to the above-discussed quantum optical many-body models. Concretely, we will consider general lattice spin-$S$ Hamiltonians with bi-linear couplings, that is,
\begin{equation}
H=-h\sum_{i}S_{i}^{z}+\sum_{i<i^{\prime}} \; \sum_{\gamma,\gamma^{\prime}\in\{+,-,z\}}J_{ii^{\prime}}^{\gamma\gamma^{\prime}}S_{i}^{\gamma}S_{i^{\prime}}^{\gamma^{\prime}}, \label{eq:Haniso_h}
\end{equation}
where $S_i^\gamma$ represents, for the $i$-th spin, its component along the $z$-direction and the spin ladder operators for $\gamma=z$ and $\gamma=+,-$, respectively. Moreover, $J_{ii^{\prime}}^{\gamma\gamma^{\prime}}$ are the coupling constants between spins at lattice sites $i$ and $i^\prime$, which may be anisotropic in the spin components and which, in cold atomic experiments, can be controlled by geometry and choice of electronic states. Finally, the homogeneous field $h$ may be generated and controlled, for example, by laser-induced level shifts. 

Our goal is to investigate the properties of the equilibrium state of these models. We assume this state to be thermal and characterized by a finite temperature $T\geq 0$, see e.g.~Ref.~\cite{sbierskiMagnetismTwodimensional2024} for the demonstration of such a thermometry approach to recent experimental correlation data from a Rydberg-array experiment \cite{chenContinuousSymmetry2023}. This is certainly an approximation for quantum optical spin systems, which are well-isolated from the environment (in the sense that they are not embedded in a solid-state matrix). Nevertheless, when preparing ground states and also excited states there is a residual entropy, manifesting in fluctuations, albeit not necessarily thermal. 

One recurrent quest in this field is to obtain phase diagrams which can be probed in experiment, e.g.~in Rydberg tweezer arrays \citep{chenContinuousSymmetry2023}. More recently, also dynamic quantities such as excitation spectra have come into reach \citep{chenSpectroscopyElementary2023}. Theoretically, these and other observables can be obtained from two-point spin correlators, see Eq.~(\ref{eq:G(=Cnu_m)}) below. For example, (second order) phase transitions are signaled by a divergence of the static correlator \cite{sandvikComputationalStudies2010}.

On the computational side, however, a large number of interaction partners per spin often causes considerable difficulties: The plain high-temperature expansion of static correlators \citep{lohmannTenthorderHightemperature2014}
becomes ineffective (see however Ref.~\cite{adelhardtMonteCarlo2024}), and approaches relying on finite simulation volumes, e.g., density matrix renormalization group (DMRG) and quantum Monte Carlo (QMC), often suffer from finite-size effects. Furthermore, these methods may be negatively affected by high dimensionality ($d>1$) or the sign-problem \citep{sandvikComputationalStudies2010}.  

As a complementary and simple method, mean-field (MF) theory
is routinely applied. It amounts to approximating the full Hamiltonian (\ref{eq:Haniso_h}) by a non-interacting trial Hamiltonian with variational parameters chosen to optimize the free energy \citep{bruusManyBodyQuantum2004}
(see also Ref.~\citep{agraFreeEnergy2006} for an insightful discussion). This procedure is often cut short by applying the mnemonically intuitive replacement 
\begin{equation}
S_{i}^{\gamma}S_{i^{\prime}}^{\gamma^{\prime}}\rightarrow S_{i}^{\gamma}\bigl\langle S_{i^{\prime}}^{\gamma^{\prime}}\bigr\rangle\!+\!\bigl\langle S_{i}^{\gamma}\bigr\rangle S_{i^{\prime}}^{\gamma^{\prime}}\!-\!\bigl\langle S_{i}^{\gamma}\bigr\rangle\bigl\langle S_{i^{\prime}}^{\gamma^{\prime}}\bigr\rangle\label{eq:OperatorMF}
\end{equation}
to the interaction term in Hamiltonian (\ref{eq:Haniso_h}), followed by a self-consistent determination of $\bigl\langle S_{i}^{\gamma}\bigr\rangle$.

In the limit of infinite connectivity, the MF approximation is quantitatively exact \citep{broutStatisticalMechanical1960,metznerCorrelatedLattice1989,defenuLongrangeInteracting2023}. Examples are all-to-all interactions (e.g.~cavity mediated in the limit of an infinite number of spins \citep{heppSuperradiantPhase1973,carolloExactness2021}) or the limit of infinite dimension. However, more realistic Hamiltonians feature a large but finite connectivity, so that the MF approximation is expected to be qualitatively correct \cite{kleinertCriticalProperties2001}. 
This regime which, as argued above, is naturally realized in many quantum optical spin systems, is the main focus of this work.  We show that the spin-spin correlator and the derived observables like phase boundaries or excitation gaps, can be well approximated by a simple and computationally inexpensive spin-diagrammatic approach. This rests on an expansion in powers of a suitably defined small parameter which varies from one problem to another. Importantly, we show that our approach also works at $T=0$ where at quantum critical points the (effective) dimensionality is increased \citep{sondhiContinuousQuantum1997}.

We build on foundations of the diagrammatic technique for general quantum spin-$S$ systems which were laid starting from the late 1960s \citep{vaksSelfconsistentField1967,vaksSpinWaves1968,stinchcombeThermalMagnetic1973,stinchcombeIsingModel1973,stinchcombeIsingModel1973a,izyumovStatisticalMechanics1988}. In Sec.~\ref{sec:Diagrammatic-technique},
we review the basic idea of the method and introduce a novel and efficient way to evaluate the diagrams to unprecedented order. For concreteness and simplicity, we specialize to specific types of spin-spin interactions in Eq.~\eqref{eq:Haniso_h}. In Sec.~\ref{sec:Results-for-Sigma} we
provide explicit expressions for the two-point functions of $h=0$ Heisenberg models and the Ising and transverse-field Ising model (TFIM), the latter at $T=0$. Throughout we stay in the symmetry unbroken regime for simplicity. In Sec.~\ref{sec:hypercubic}
we benchmark our method on the hypercubic-lattice with nearest-neighbor interactions. Finally, in Sec.~\ref{sec:applications}, we apply our method to two particular quantum optical many-body systems with $S=1/2$, both on the square lattice: A power-law interacting ferromagnetic (FM) Heisenberg model at $T>0$ \citep{zhaoFinitetemperatureCritical2023}
and the Dicke-Ising model at $T=0$ \citep{gelhausenQuantumopticalMagnets2016, schellenbergerAlmostEverything2024a}.
We conclude in Sec.~\ref{sec:conclusion}.
%%%%%%%%%%%%%%%%%%%%%%%%%%%%%%%%%%%%%%%%%%%%%%%%%%%%%%%%%%%%%%%
%%%%%%%%%%%%%%%%%%%%%%%%%%%%%%%%%%%%%%%%%%%%%%%%%%%%%%%%%%%%%%%
\section{Diagrammatic technique for quantum spins via kernel functions\label{sec:Diagrammatic-technique}}

%%%%%%%%%%%%%%%%%%%%%%%%%%%%%%%%%%%%%%%%%%%%%%%%%%%%%%%%%%%%%%%
\subsection{Models and perturbative expansion}
\label{subsec:models}

We consider SU(2) quantum spin-$S$ operators $\mathbf{S}_{j}=(S_{j}^{x},S_{j}^{y},S_{j}^{z})^{\mathrm{T}}$
on a lattice with $N$ sites labeled by index $j$. The operators
obey the spin algebra 
\begin{equation}
\left[S_{j_{1}}^{\alpha_{1}},S_{j_{2}}^{\alpha_{2}}\right]=i\delta_{j_{1}j_{2}}\epsilon^{\alpha_{1}\alpha_{2}\alpha_{3}}S_{j_{1}}^{\alpha_{3}}\label{eq:spinCommutator}
\end{equation}
with $\alpha_{1,2,3}\in\{x,y,z\}$ and the spin-length operator constraint
$\mathbf{S}_{j}\cdot\mathbf{S}_{j}=S(S+1)$. The general bi-linear spin Hamiltonian with homogeneous magnetic field $h$ in z-direction is given
in Eq.~(\ref{eq:Haniso_h}) where $S^{\pm}=(S^{x}\pm iS^{y})/\sqrt{2}$ 
are the spin ladder operators. In the following, it is understood that $\alpha \in \{ x,y,z \}$ while $\gamma \in \{+,-,z\}$.

In the following, we specialize Eq.~(\ref{eq:Haniso_h}) to three important model classes defined by
particular combinations of $h$ and the $3\times 3$ matrix $J_{ii^{\prime}}^{\gamma\gamma^{\prime}}$ in flavor space. This specialization is not required by methodological restrictions but was chosen to keep the notation and diagrammatic complexity at a level suitable for presentation. Finally, in Sec.~\ref{subsec:Dicke-Ising-model:-Ground-state} we consider a model with a more complicated structure.

The three simple choices correspond
to the following standard spin models: (i) The Ising
model (which is classical and treated mainly for reference)
\begin{equation}
H=\sum_{i<i^{\prime}}J_{ii^{\prime}}S_{i}^{z}S_{i^{\prime}}^{z}\label{eq:Ising}
\end{equation}
is obtained from Eq.~(\ref{eq:Haniso_h}) by setting $h=0$ and $J_{ii^{\prime}}^{\gamma\gamma^{\prime}}=\delta_{\gamma,z}\delta_{\gamma^{\prime}z}J_{ii^{\prime}}$
. (ii) The transverse field Ising model (TFIM)
\begin{equation}
H=\sum_{i<i^{\prime}}J_{ii^{\prime}}S_{i}^{x}S_{i^{\prime}}^{x}-h\sum_{i}S_{i}^{z}\label{eq:TFIM}
\end{equation}
corresponds to $J_{ii^{\prime}}^{\gamma\gamma^{\prime}}=\frac{1}{2}\left(1-\delta_{\gamma,z}\right)\left(1-\delta_{\gamma^{\prime},z}\right)J_{ii^{\prime}}$.
Note that Ising models are usually defined only for $S=1/2$. (iii) The
Heisenberg model 
\begin{equation}
H=\sum_{i<i^{\prime}}J_{ii^{\prime}}\mathbf{S}_{i}\cdot\mathbf{S}_{i^{\prime}}\label{eq:Heisenb}
\end{equation}
results from $h=0$ and $J_{ii^{\prime}}^{\gamma\gamma^{\prime}}=\delta_{\bar{\gamma},\gamma^{\prime}}J_{ii^{\prime}}$
where we define $\bar{\gamma}$ by $\bar{+}=-$, $\bar{-}=+$ and
$\bar{z}=z$. We keep $S$ general and assume
vanishing onsite-coupling $J_{ii}=0$ for simplicity.

In the remainder of this work, our computational focus will be on the Matsubara spin-spin correlation function,
\begin{equation}
G_{jj^{\prime}}^{\alpha\alpha}(i\nu_{m})=T\int_{0}^{\beta}\mathrm{d}\tau\mathrm{d}\tau^{\prime}\,e^{i\nu_{m}(\tau-\tau^{\prime})}\,G_{jj^{\prime}}^{\alpha\alpha}(\tau,\tau^{\prime}),\label{eq:G(=Cnu_m)}
\end{equation}
where $\tau$ is (imaginary) time, $\beta=1/T$ is the inverse temperature,
$\nu_{m}=2\pi Tm$ with $m\in\mathbb{Z}$ a (bosonic) Matsubara frequency
and
\begin{equation}
G_{j j^\prime}^{\alpha\alpha}(\tau,\tau^{\prime})
=\left\langle \mathcal{T}S_{j}^{\alpha}(\tau)S_{j^{\prime}}^{\alpha}(\tau^{\prime})\right\rangle =G_{j j^\prime}^{\alpha\alpha}(\tau-\tau^{\prime}) \label{eq:G_Matsubara_tau}
\end{equation}
the time-ordered (``$\mathcal{T}$'') thermal correlation
function \citep{izyumovStatisticalMechanics1988} which only depends on the time difference. Time ordering for spin operators is defined via
\begin{equation}
\mathcal{T}S_{j}^{\alpha}(\tau)S_{j^{\prime}}^{\alpha^{\prime}}(\tau^{\prime})=\begin{cases}
S_{j}^{\alpha}(\tau)S_{j^{\prime}}^{\alpha^{\prime}}(\tau^{\prime}) & :\tau>\tau^{\prime},\\
S_{j^{\prime}}^{\alpha^{\prime}}(\tau^{\prime})S_{j}^{\alpha}(\tau) & :\tau^{\prime}>\tau,
\end{cases}
\end{equation}
and operators in imaginary-time Heisenberg picture are written as $S_{j}^{\alpha}(\tau)= e^{H\tau}S_{j}^{\alpha}e^{-H\tau}$.
Thermal averages are defined as $\left\langle ...\right\rangle =\mathcal{Z}^{-1}\mathrm{Tr}\left[e^{-\beta H}...\right]$
where $\mathcal{Z}=\mathrm{Tr}e^{-\beta H}$ is the partition function.

While dynamic observables
are encoded in the analytically continued version of Eq.~\eqref{eq:G(=Cnu_m)}, $G_{j j^\prime}^{\alpha\alpha}(i\nu_{m}\rightarrow\nu\pm i\eta)$, we will be mainly concerned with the detection of magnetic phase transitions of second-order (continuous) nature.
Such a transition towards an ordered phase (with $\bigl\langle S_{i}^{\alpha}\bigr\rangle\neq0$,
say) can be conveniently detected coming from the paramagnetic side
without spontaneous symmetry breaking, see the black arrow in Fig.~\ref{fig:PT-Heisenberg}(a). 
According to isothermal linear response theory, it is signaled by
a divergent spatial Fourier transform (at ordering wavevector $\mathbf{Q}$)
of the static Matsubara spin-spin correlator [Eq.~\eqref{eq:G(=Cnu_m)} with $\nu_{m}=0$] also known as spin susceptibility \cite{sandvikComputationalStudies2010}. Note that around first-order (discontinuous) transitions, where the order parameter jumps, the susceptibility is not defined. The analysis or detection of first-order transitions is beyond the scope of this work.

Given these considerations, in the following we will focus on this symmetry-unbroken phase and consider the Matsubara correlator \eqref{eq:G(=Cnu_m)} for $\alpha$ for which $\bigl\langle S_{i}^{\alpha}\bigr\rangle=0$. Among other simplifications on the diagrammatic side to be discussed below, this choice ensures the equality of connected and disconnected two-point correlators.

Next, we review the diagrammatic series expansion of $G_{j j^\prime}^{\alpha\alpha}(i\nu_{m})$
in exchange interaction $J$, originally developed for the case of quantum spins in Refs.~\citep{vaksSelfconsistentField1967,vaksSpinWaves1968,stinchcombeThermalMagnetic1973,stinchcombeIsingModel1973,stinchcombeIsingModel1973a}
and summarized in Ref.~\citep{izyumovStatisticalMechanics1988}. Earlier work on the (classical) Ising model can be found in Refs.~\cite{broutStatisticalMechanical1960,horwitzDiagrammaticExpansion1961,englertLinkedCluster1963}.
As usual in perturbation theory \citep{bruusManyBodyQuantum2004},
we start by splitting the Hamiltonian in interacting and non-interacting
parts. For the general spin Hamiltonian (\ref{eq:Haniso_h}), we thus
set $H=H_{0,h}-V$ where
\begin{equation}
H_{0,h}\!=\!-h\!\sum_{i}S_{i}^{z},\;\;\; V\!=\!-\!\!\sum_{i<i^{\prime}}\sum_{\gamma,\gamma^{\prime}\in\{+,-,z\}}\!\!\!\!\!\!\!J_{ii^{\prime}}^{\gamma\gamma^{\prime}}S_{i}^{\gamma}S_{i^{\prime}}^{\gamma^{\prime}}.\label{eq:H0h_V}
\end{equation}
The formal series expansion in $V\sim J$ reads \citep{bruusManyBodyQuantum2004}
\begin{align}
G_{jj^{\prime}}^{\gamma\gamma^{\prime}}(\tau,\tau^{\prime})\! & =\!\sum_{n=0}^{\infty}\frac{1}{n!}\!\int_{0}^{\beta}\!\mathrm{d}\tau_{1}...\mathrm{d}\tau_{n}\label{eq:G_series}\\
 & \times\left\langle \!\mathcal{T}V(\tau_{1})...V(\tau_{n})S_{j}^{\gamma}(\tau)S_{j^{\prime}}^{\gamma^{\prime}}(\tau^{\prime})\!\right\rangle _{0,h,V \mkern-4mu -\mkern-1.5mu c}.\nonumber 
\end{align}
Here, the averages and time evolution of operators are governed only by $H_{0,h}$ (we avoid a new symbol since no confusion is
possible from here on). For the correlator $G$, we changed to flavor indices $\gamma,\gamma^{\prime}\in\{+,-,z\}$
adapted to the U(1) symmetry of $H_{0,h}$. This will prove
convenient below. 
The subscript $V \!\! - \!\! c$ for the averages on the right-hand
side of Eq.~(\ref{eq:G_series}) excludes vacuum contributions
\citep{rossiDeterminantDiagrammatic2017} where in a diagrammatic interpretation one or more $V(\tau)$ are
not connected to the external indices after performing the average. We refer to App.~\ref{app:V-con} for a rigorous definition. 
Note that a formula analogous to Eq.~(\ref{eq:G_series})
applies for any time-ordered product of Heisenberg-picture operators and in particular also for the local magnetization $\text{\ensuremath{\bigl\langle}\ensuremath{\mathcal{T}S_{j}^{\alpha}}(\ensuremath{\tau})\ensuremath{\bigr\rangle}}=\bigl\langle S_{j}^{\alpha}\bigr\rangle$. However, as mentioned above, in this work we will focus only on two-point functions.

Finally we perform the $\tau,\tau^{\prime}$-integral
over Eq.~(\ref{eq:G_series}) to obtain the Matsubara correlator (\ref{eq:G(=Cnu_m)})
and also specialize to a fixed expansion order $n$ indicated by a superscript, 
\begin{eqnarray}
G_{jj^{\prime}}^{\gamma\gamma^{\prime}(n)}(i\nu_{m}) & = & T\,\frac{1}{n!}\int_{0}^{\beta}e^{i\nu_{m}(\tau-\tau^{\prime})}\mathrm{d}\tau_{1}...\mathrm{d}\tau_{n}\mathrm{d}\tau\mathrm{d}\tau^{\prime} \label{eq:G(nu_m)_series}\\
& \times &
\left\langle \mathcal{T}V(\tau_{1})...V(\tau_{n})S_{j}^{\gamma}(\tau)S_{j^{\prime}}^{\gamma^{\prime}}(\tau^{\prime})\right\rangle _{0,h,V \mkern-4mu -\mkern-1.5mu c}.\nonumber
\end{eqnarray}

%%%%%%%%%%%%%%%%%%%%%%%%%%%%%%%%%%%%%%%%%%%%%%%%%%%%%%%%%%%%%%%%%%%%%
\subsection{Diagrams, J-reducibility and MF approximation}
\label{subsec:Diagrams,1JI,MF}

We proceed with the evaluation of Eq.~\eqref{eq:G(nu_m)_series} by a diagrammatic approach \citep{izyumovStatisticalMechanics1988}. This is conceptually simple, physically transparent
and allows for various subsequent approximation and resummation schemes
one of which will be essential later. In the following,
we review the diagrammatic method, the concept of J-reducibility
and its relation to the MF approximation (\ref{eq:OperatorMF}). In the next subsection, we will propose a novel evaluation scheme of the diagrams that provides analytical expressions at high orders $J^n$.

\begin{figure}
\begin{centering}
\includegraphics{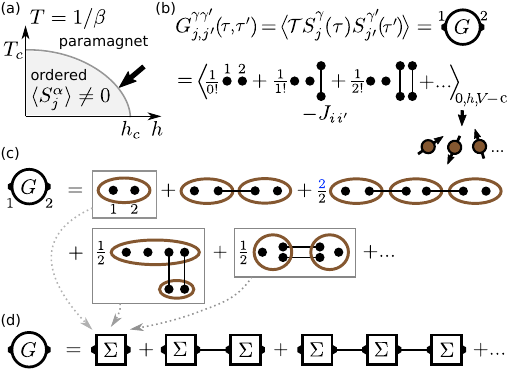}
\par\end{centering}
\caption{\label{fig:PT-Heisenberg}(a) Schematic of a typical phase diagram
for Hamiltonian (\ref{eq:Haniso_h}). In this work we approach the
magnetic ordering transition from the paramagnetic phase. (b) Diagrammatic
representation of the expansion for the spin-spin correlator
$G$ in the exchange interaction $-J$ (lines), the orders $n=0,1,2$ are shown
explicitly. (c) Diagrammatic representation of the $V$-connected
correlators in (b) in terms of ordinary connected free-spin correlators
(brown ellipses). The 1-J-irreducible diagrams of order $J^0$ and $J^2$ are marked with boxes,
the infinite sum of all these diagrams constitutes $\Sigma$, the 1-J-irreducible part of $G$.
(d) The Larkin equation re-combines $\Sigma$ and $J$ to form $G$.}
\end{figure}

The diagrammatic formulation starts from the graphical representation of Eq.~\eqref{eq:G_series} in Fig.~\ref{fig:PT-Heisenberg}(b).
The dots represent spin operators and for now we do not perform
the Fourier transform to Matsubara frequency. The
two external spin operators carry multi-indices $1=(\gamma,j,\tau)$ and
$2=(\gamma^{\prime},j^{\prime},\tau^{\prime})$. All other (internal)
spin operators coming from $V$ appear pairwise connected by interaction
lines representing $-J_{i_{k}i_{k}^{\prime}}$, $k=1,2,..,n$. Flavors,
times and sites of the internal spin operators are summed (or integrated)
over and the leading factor $1/n!$ is written explicitly. Interaction
lines connect spin operators at the same time (unless retarded,
c.f.~Sec.~\ref{subsec:Dicke-Ising-model:-Ground-state}). 

Next, the $V$-connected free spin average in Eq.~\eqref{eq:G_series} is taken by co-localizing the
$2+2n$ spin operators (dots) in any possible way involving blocks
of $m=2,3,4,...$ spins shown as brown ellipses, see Fig.~\ref{fig:PT-Heisenberg}(c) \cite{izyumovStatisticalMechanics1988}. These blocks represent connected equal-site
time-ordered free spin correlators,
\begin{equation}
\left\langle \mathcal{T}S_{i}^{\gamma_{1}}(\tau_{1})...S_{i}^{\gamma_{m}}(\tau_{m})\right\rangle _{0,c,h}=G_{0,c,h}^{\gamma_{1}...\gamma_{m}}\left(\tau_{1},...,\tau_{m}\right),\label{eq:G0con}
\end{equation}
which, for the particular non-interacting Hamiltonian $H_{0,h}$ that
we consider in Eq.~\eqref{eq:H0h_V}, do not depend on the site index $i$. Connected spin correlators (with subscript $c$) are defined in analogy to
their $V$-connected counterparts, see App.~\ref{app:V-con}. Due to our choice of $J_{ii}=0$, interaction lines cannot start and end at the same block. 
In Fig.~\ref{fig:PT-Heisenberg}(c), the absence of blocks of size $m=1$ representing single-spin averages (magnetization) is due to our current focus on the three models in question and on $G^{\alpha \alpha}$ for which $\bigl\langle S_{i}^{\alpha}\bigr\rangle=0$ in the symmetry-unbroken regime. The presence of blocks of order $m>2$
signals the absence of Wick's theorem \citep{bruusManyBodyQuantum2004}
for spin operators, ultimately rooted in the operator-valued right-hand
side of the commutation relation (\ref{eq:spinCommutator}) which
differs from the canonical case of bosonic creation and annihilation operators.

The connected time-ordered free spin correlators of Eq.~\eqref{eq:G0con}, which are central diagrammatic building blocks will be discussed below in generality. For now, we mention the important special case that arises if all $m$ flavor indices are chosen as $z$ such that all involved operators $S_i^z(\tau_n)=S_i^z$ commute. These connected z-spin correlators are then time-independent and given via the order-$(m-1)$ derivative of the Brillouin
function $b_{c}(y)$ which describes the magnetization of a free spin $\left\langle S^z \right\rangle$ dependent on $y=\beta h$ \citep{izyumovStatisticalMechanics1988},
\begin{eqnarray}
G_{0,c,h}^{z...z}\left(\tau_{1},...,\tau_{m}\right) & = & b_{c}^{(m-1)}(\beta h),\label{eq:Gz...z,0,con}
\end{eqnarray}
\begin{equation}
b_{c}(y)=(S+\frac{1}{2})\coth\left[(S+\frac{1}{2})y\right]-\frac{1}{2}\coth\frac{y}{2}.\label{eq:bcy}
\end{equation}
For $h\rightarrow0$ we abbreviate $b_{c}^{(m-1)}(0)\equiv b_{c,m-1}$,
e.g.~$b_{c,1}=S(S+1)/3$ is the free spin (static) Curie susceptibility that will become important in the following discussion.

Having explained the building blocks of the diagrams in Fig.~\ref{fig:PT-Heisenberg}(c), we now turn their topology, i.e.~the particular choice of blocks and their connection.
Topological multiplicity factors $\Lambda$ (denoted in blue) appear if the same diagram topology can be arrived at by distributing the dots in Fig.~\ref{fig:PT-Heisenberg}(b) to the given set of blocks
in multiple ways, see e.g.~the last diagram in the first line of
Fig.~\ref{fig:PT-Heisenberg}(c). Diagrams that only differ by a
rearrangement of dots \emph{within} blocks are not topologically different. Formally, the multiplicity factor can be computed via the number of elements in the automorphism group $\mathcal{A}_{x}$ of a diagram $x$ of order $J^n$ : $\Lambda = n! / \mathrm{ord}(\mathcal{A}_{x})$.

The diagrams in Fig.~\ref{fig:PT-Heisenberg}(c) can be classified
as either 1-J reducible or 1-J irreducible (1JI). In 1-J reducible
diagrams it is possible to separate the external
spin operators (single dots) by cutting a single interaction line. If this is not possible, the diagram is 1JI. The infinite sum of all 1JI diagrams is denoted by $\Sigma$. For examples, the contributions to $\Sigma$ of order $J^{0}$
and $J^{2}$ are indicated with boxes in Fig.~\ref{fig:PT-Heisenberg}(c). 

All diagrammatic contributions to the spin correlator $G$ that are 1-J reducible can be represented by
1JI diagrams connected by one or more interaction lines,
see Fig.~\ref{fig:PT-Heisenberg}(d). This diagrammatic expression can be summed exactly and yields the Larkin equation \citep{vaksSelfconsistentField1967,izyumovStatisticalMechanics1988}.
In frequency space, it reads 
\begin{equation}
G(i\nu_{m})=\left(1+\Sigma(i\nu_{m})\cdot J\right)^{-1}\cdot\Sigma(i\nu_{m}).\label{eq:Larkin}
\end{equation}
Here all objects are considered $N\times N$ matrices with two site indices,
c.f.~Eq.~(\ref{eq:G(=Cnu_m)}) and for our purposes the flavors
will all be set to a fixed $\alpha$ with $\alpha\in\{x,y,z\}$ for
simplicity. Note the similarity of Eq.~(\ref{eq:Larkin}) to Dyson's
equation for canonical bosonic or fermionic systems \citep{bruusManyBodyQuantum2004}
which however rests on the concept of cutting free propagator lines (and
not interaction lines). 

\begin{figure}[t]
\begin{centering}
\includegraphics{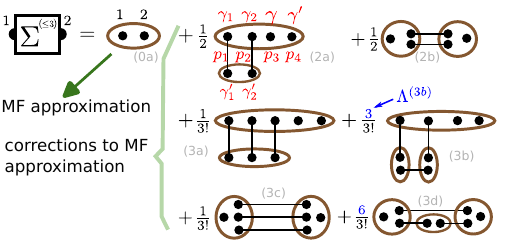}
\par\end{centering}
\caption{\label{fig:Sigma023-Heisenberg} Diagrams for $\Sigma$
up to order $J^{3}$ for the situation described in the text where magnetization is not relevant. The blue multiplicity
factors $\Lambda^{(nx)}$ denote the number of ways to arrive at a given diagram starting from the expansion in Fig.~\ref{fig:PT-Heisenberg}(b),
e.g.~in diagram (3b) there are three choices to pick the bottom line
from the three lines available at third order perturbation theory.
The two vertical lines are equivalent and do not enhance the multiplicity
as an exchange can be compensated by a permutation of sites within
the top ellipse.}
\end{figure}

Given the above diagrammatic rules, it is straightforward to write
down the 1JI diagrams for $\Sigma$ at order $n$ indicated by $\Sigma^{(n)}$. The diagrams for $n\leq3$ which include the boxed diagrams of Fig.~\ref{fig:PT-Heisenberg}(b) are shown in Fig.~\ref{fig:Sigma023-Heisenberg}.
We label the different diagram topologies at order $n$ with an additional index $x=a,b,c,...$. Each diagram topology is uniquely identified by $(nx)$.
The diagrams for $\Sigma^{(4)}$ have not been systematically collected
so far in literature and are given in Fig.~\ref{fig:=CSigma(4)}
of App.~\ref{app:DiagramsSigma4}. Note that diagrams that combine the two external operators together with one internal operator in a $m=3$ block are 1JI by our definition. However, for the particular setup and observables that we focus on in this work (see above), these diagrams vanish.

We conclude this subsection by linking the diagrammatic expansion
to the MF approximation for $G^{\alpha\alpha}$. Quite simply, the latter is obtained
by using only the lowest-order approximation for $\Sigma$ (green arrow in Fig.~\ref{fig:Sigma023-Heisenberg})
\begin{equation}
\Sigma^{\alpha\alpha}\rightarrow\Sigma^{\alpha\alpha(0)}=G_{0,c,h}^{\alpha\alpha}.\label{eq:MF-approx}
\end{equation}
For $\alpha=z$, this evaluates to $b_{c}^{(1)}(\beta h)$, see Eq.~\eqref{eq:Gz...z,0,con}.
We also assumed $\left\langle S_{i}^{\alpha}\right\rangle =0$, otherwise
$h$ in Eq.~\eqref{eq:MF-approx} would need to be replaced by the Weiss
field, see Ref.~\citep{izyumovStatisticalMechanics1988} and Sec.~\ref{subsec:Dicke-Ising-model:-Ground-state}. The easiest way to
see the correspondence between Eq.~(\ref{eq:MF-approx}) and the
MF approximation is to recall that the latter is exact for
$N\rightarrow\infty$ if each spin interacts with all other spins
in the system which requires to redefine $J\rightarrow J/N$. For $G$ and in the limit $N\rightarrow\infty$, this means that string-like
diagrams as in the first line of Fig.~\ref{fig:PT-Heisenberg}(c)
are the only finite contribution. This follows since the
site-indices of the internal ($m=2$-)blocks can be summed freely over all available
sites only in string-like diagrams while being more restricted in diagrams that contain loops. Finally, the string-like diagrams are exactly the diagrams that
are produced by the replacement $\Sigma\rightarrow\Sigma^{(0)}$
in the Larkin equation of Fig.~\ref{fig:PT-Heisenberg}(d).
It follows that contributions $\Sigma^{(n)}$ with $n>0$ represent corrections to the MF results, see Figs.~\ref{fig:JConnectivityOverview}
and \ref{fig:Sigma023-Heisenberg}.
 
%%%%%%%%%%%%%%%%%%%%%%%%%%%%%%%%%%%%%%%%%%%%%%%%%%%%%%%%%%%%%%%%%
\subsection{Diagram evaluation via kernel functions}
\label{subsec:diagEval}

The remaining task is to calculate the analytical expressions
encoded in the particular diagrams $\Sigma_{jj^{\prime}}^{\gamma\gamma^{\prime}(nx)}(i\nu_{m})$ which represent the different contributions of the right-hand side of Eq.~\eqref{eq:G(nu_m)_series}. Two observations on the simple nature of the chosen models (Ising, TFIM and Heisenberg, c.f.~Sec.~\ref{subsec:models}) are helpful but by no means crucial for progress: (a) The values for the blocks [brown ellipses, see Eq.~\eqref{eq:G0con}] are site-independent and (b) the coupling between any two blocks is characterized by a single number $J_{ii'}$ (unlike for, say, XXZ interactions). Hence we can combine all $J_{ii^\prime}$ [stemming from the insertions of $V$, c.f.~Eq.\eqref{eq:H0h_V}] along with the site sums in a geometry factor
$t^{(nx)}_{jj^{\prime}}\left[J\right]\sim J^{n}$,
\begin{equation}
\Sigma_{jj^{\prime}}^{\gamma\gamma^{\prime}(nx)}(i\nu_{m})\equiv t^{(nx)}_{jj^{\prime}}\!\left[J\right]\cdot\sigma^{\gamma\gamma^{\prime}(nx)}(i\nu_{m}).\label{eq:Sigma=t*sigma}
\end{equation}
This geometry factor captures the dependence of the particular diagram $(nx)$ on $j,j^\prime$ and the underlying $N\times N$ coupling matrix $J_{ii^\prime}$. For example, with regard to Fig.~\ref{fig:Sigma023-Heisenberg}, the geometry factor for diagram $(0a)$ is simply $\delta_{jj^\prime}$ whereas for $(2a)$ and $(2b)$ we read off
\begin{equation}
    t^{(2a)}_{jj^\prime}=\delta_{jj^\prime} \sum_i J_{ji} J_{ij}, \;\;\;\;
    t^{(2b)}_{jj^\prime}=(J_{jj^\prime})^2.
\end{equation}
The geometry factors for all diagrams of order $n=0,2,3$ are shown in the second column in Tab.~\ref{tab:t023}. In the third column, we provide the
momentum space representation for translation invariant lattices with
spatial Fourier transform defined by
\begin{equation}
J_{\mathbf{k}}=\sum_{\mathbf{r}_{j}}e^{-i\mathbf{k}\cdot\mathbf{r}_{j}}J_{0,\mathbf{r}_{j}},\label{eq:k-FT}
\end{equation}
and analogous for $G_{\mathbf{k}}$ and $\Sigma_{\mathbf{k}}$. In
the fourth column we specialize to the infinite ($N\rightarrow\infty$)
nearest-neighbor hypercubic lattice model in $d$ spatial dimensions,
see the caption of Tab.~\ref{tab:t023} for more details. Geometry factors for the fourth-order diagrams $\Sigma^{(4)}$ are summarized in Tab.~\ref{tab:t4} of App.~\ref{app:DiagramsSigma4}.
\begin{table}
\begin{centering}
\begin{tabular}{c|l|l|l}
top. & r-space $t_{jj^\prime}^{(nx)}$ & k-space $t_{\mathbf{k}}^{(nx)}$ & nn-hyp. $\tilde{t}_{\mathbf{k}}^{(nx)}$\tabularnewline
\hline 
(0a) & $\delta_{jj^\prime}$ & $1$ & $1$\tabularnewline
\hline 
(2a) & $\delta_{jj^\prime}\left[J\cdot J\right]_{jj^\prime}$ & $\int_{\mathbf{q}}J_{\mathbf{q}}^{2}$ & $J_{1}\tilde{J}_{\mathbf{0}}$\tabularnewline
(2b) & $J_{jj^\prime}^{2}$ & $\int_{\mathbf{q}}J_{\mathbf{q}-\mathbf{k}}J_{\mathbf{q}}$ & $J_{1}\tilde{J}_{\mathbf{k}}$\tabularnewline
\hline 
(3a) & $\delta_{jj^\prime}\sum_{i}J_{ji}^{3}$ & $\int_{\mathbf{p},\mathbf{q}}J_{\mathbf{p}}J_{\mathbf{q}}J_{\mathbf{p}+\mathbf{q}}$ & $J_{1}^{2}\tilde{J}_{\mathbf{0}}$\tabularnewline
(3b) & $\delta_{jj^\prime}\left[J\cdot J\cdot J\right]_{jj}$ & $\int_{\mathbf{q}}J_{\mathbf{q}}^{3}$ & $0$\tabularnewline
(3c) & $J_{jj^\prime}^{3}$ & $\int_{\mathbf{q},\mathbf{p}}J_{\mathbf{p}-\mathbf{k}}J_{\mathbf{q}}J_{\mathbf{p}+\mathbf{q}}$ & $J_{1}^{2}\tilde{J}_{\mathbf{k}}$\tabularnewline
(3d) & $J_{jj^\prime}\left[J\cdot J\right]_{jj^\prime}$ & $\int_{\mathbf{q}}J_{\mathbf{k}-\mathbf{q}}J_{\mathbf{q}}^{2}$ & $0$\tabularnewline
\end{tabular}
\par\end{centering}
\caption{\label{tab:t023}Geometry factors $t^{(nx)}\left[J\right]$ for $\Sigma$-diagrams
of order $n=0,2,3$ and all topologies (for the case $n=4$ see Tab.~\ref{tab:t4}).
Results in the second column are given in real space {[}where $J_{jj^\prime}^{2}$
is understood as $(J_{jj^\prime})^{2}$ etc.{]}. The third column
shows the same results in momentum space (where we abbreviate $\int_{\mathbf{q}}=1/N\sum_{\mathbf{q}}$).
In the fourth column, momentum space results are specialized for an
infinite $d$-dimensional hypercubic lattice with a lattice constant
of unity and nearest-neighbor coupling $J_{1}$ where $\tilde{J}_{\mathbf{k}}\equiv2J_{1}\sum_{\mu=1}^{d}\cos k_{\mu}$
and accordingly $\tilde{J}_{\mathbf{0}}=2dJ_{1}$.}
\end{table}

Beyond the frequency independent geometry factor, the remaining contribution in Eq.~\eqref{eq:Sigma=t*sigma} denoted as $\sigma^{(nx)}(i\nu_m)$ only depends on the type of spin model (here: Ising, TFIM, Heisenberg). We define it to include the topological multiplicity factor $\Lambda^{(nx)}$ of the diagram $(nx)$.
In principle, obtaining $\sigma^{(nx)}$ is straightforward in frequency space \citep{izyumovStatisticalMechanics1988}
where interaction lines proportional to $\delta_{\nu_{l}+\nu_{l^{\prime}}}$
connect the blocks which are temporal Fourier transforms of Eq.~\eqref{eq:G0con},
$G_{0,c,h}^{\gamma_{1}...\gamma_{m}}(\omega_{1},...,\omega_{m-1})$.
The latter, however, are difficult to calculate in general beyond the case $\gamma_1=...=\gamma_m=z$ [see Eq.~\eqref{eq:Gz...z,0,con}], especially for
$h=0$ where a distinction of cases regarding various frequency combinations is required. Despite recent algorithmic advances \citep{gollSpinFunctional2019,halbingerSpectralRepresentation2023}
the general $G_{0,c}^{\gamma_{1}...\gamma_{m}}(\omega_{1},...,\omega_{m-1})$ (which beyond their appearance in perturbation theory lack physical significance for large $m$)
are currently known analytically up to order $m=4$ \citep{izyumovStatisticalMechanics1988}. This
would only suffice to compute $\Sigma$ to order $J^2$.
On the other hand, without the temporal Fourier transform and for a fixed order of times, the blocks representing $\left\langle \mathcal{T}S^{\gamma_{1}}(\tau_{1})...S^{\gamma_{m}}(\tau_{m})\right\rangle _{0,c,h}$ [c.f.~Eq.~\eqref{eq:G0con}] simplify to standard connected free-spin equal-time averages
$\left\langle S^{\tilde{\gamma}_{1}}...S^{\tilde{\gamma}_{m}}\right\rangle _{0,c,h}$
which can be efficiently computed for general $S$, see Sec.~6.2.~in
Ref.~\citep{halbingerSpectralRepresentation2023}.

To make progress, the crucial reformulation detailed in App.~\ref{app:KernelFunction} expresses $\sigma^{(nx)}$ such that only the simple $\left\langle S^{\tilde{\gamma}_{1}}...S^{\tilde{\gamma}_{m}}\right\rangle _{0,c,h}$ are required. All the remaining complexity is encapsulated in the universal (model independent) kernel functions $K_{n+2}(\Omega_1,...,\Omega_{n+2})$ of Ref.~\cite{halbingerSpectralRepresentation2023}. The latter originally appeared in the context of Fourier transforms of time-ordered correlation functions. The key point is that an $n$-th order
perturbative expression as in Eq.~\eqref{eq:G(nu_m)_series} can be understood 
as a Fourier transform of a time-ordered correlation function of order
$n+2$ with $n$ frequencies set to zero.  

The translation of diagram $(nx)$ to an expression $\sigma^{(nx)}$
is straightforward. To avoid unnecessary complicated notation, we
just give two examples: For diagram $(2a)$, with the labeling of internal times and flavors in Fig.~\ref{fig:Sigma023-Heisenberg} (red), one obtains
\begin{widetext}
\begin{eqnarray}
\sigma^{\gamma\gamma^{\prime}(2a)}(i\nu_{m}) &=&\Lambda^{(2a)}\frac{(-1)^{n}}{n!}\sum_{\gamma_{1,..,n}^{(\prime)}}\; \sum_{p\in S_{n+2}} K_{n+2}\left(\mathcal{P}\left\{ \Omega_{1}(p_{1}),...,\Omega_{n+2}(p_{n+2})\right\} \right) \label{eq:(2a)} \\
&\times& \left\langle \mathcal{P}S^{\gamma_{1}}(p_{1})S^{\gamma_{2}}(p_{2})S^{\gamma}(p_{3})S^{\gamma^{\prime}}(p_{4})\right\rangle _{0,c,h} 
\left\langle \mathcal{P}S^{\gamma_{1}^{\prime}}(p_{1})S^{\gamma_{2}^{\prime}}(p_{2})\right\rangle _{0,c,h}, \nonumber
\end{eqnarray}
where $\Lambda^{(2a)}=1$ and
$n=2$. As another example, diagram (3b) with $\Lambda^{(3b)}=3$ and $n=3$ is
\begin{eqnarray}
\sigma^{\gamma\gamma^{\prime}(3b)}(i\nu_{m})&=&\Lambda^{(3b)}\frac{(-1)^{n}}{n!}\sum_{\gamma_{1,...,n}^{(\prime)}} \; \sum_{p\in S_{n+2}}  
K_{n+2}\left(\mathcal{P}\left\{ \Omega_{1}(p_{1}),...,\Omega_{n+2}(p_{n+2})\right\} \right) \label{eq:(3b)} \\
& \times &
\left\langle \mathcal{P}S^{\gamma_{1}}(p_{1})S^{\gamma_{2}}(p_{2})S^{\gamma}(p_{4})S^{\gamma^{\prime}}(p_{5})\right\rangle _{0,c,h} 
 \left\langle \mathcal{P}S^{\gamma_{1}^{\prime}}(p_{1})S^{\gamma_{3}}(p_{3})\right\rangle _{0,c,h}\left\langle \mathcal{P}S^{\gamma_{2}^{\prime}}(p_{2})S^{\gamma_{3}^{\prime}}(p_{3})\right\rangle _{0,c,h}.\nonumber 
\end{eqnarray}
The internal flavor sums over $\gamma_{1,...,n}^{(\prime)}$ depend on the model, e.g.~for the Heisenberg case these sums are over $\{+,-,z\}$ while restricted to $\{z\}$ in the Ising case. The second sum is over the $(n+2)!$ permutations $S_{n+2}$ that determine the ordering of both the argument list of $K_{n+2}$ and the spin operators in the equal-time averages. This is accomplished by the index
ordering operator $\mathcal{P}$. This operator applies to operator
strings and argument lists alike. It acts like time-ordering, but
for discrete indices $(1),(2),...,(n+2)$ that - unlike imaginary
time arguments - do not affect the operator, for example $\mathcal{P}S^+(1)S^-(3)S^z(2)=S^-S^zS^+$.
Finally, the $\Omega$-list is given by \begin{eqnarray}
\left\{ \Omega_{1},...,\Omega_{n},\Omega_{n+1},\Omega_{n+2}\right\}  & = & \left\{ -(\gamma_{1}+\gamma_{1}^{\prime})h,...,-(\gamma_{n}+\gamma_{n}^{\prime})h,i\nu_{m}-\gamma h,-i\nu_{m}-\gamma^{\prime}h\right\},\label{eq:=COmega_list}
\end{eqnarray}
where the following replacement rule for flavor labels is understood: $\{z,+,-\}\rightarrow\{0,+1,-1\}$.
\end{widetext}

Closed-form expressions for general kernel functions $K_{k}(\Omega_{1},...,\Omega_{k})$
can be found in Ref.~\citep{halbingerSpectralRepresentation2023}.
The expressions grow in complexity with $k$, however, for the current application with the specific form of the $\Omega$-list in Eq.~(\ref{eq:=COmega_list})
and our focus on either $h=0$ (Ising and Heisenberg models) or $T\rightarrow0$
(for the TFIM), the kernel functions simplify considerably. The resulting
expressions are given in App.~\ref{app:Special-Kernel-functions}.

Expressions analogous to Eqns.~\eqref{eq:(2a)} and \eqref{eq:(3b)} can be straightforwardly written for all diagram topologies and should be evaluated via computer algebra for efficiency. In the next
section, we provide the expressions for $\sigma^{\gamma\gamma^{\prime}(nx)}(i\nu_{m})$
obtained in this way. Finally, the sum over all topologies yields $\Sigma^{(n)}$
at order $J^{n}$, 
\begin{equation}
\Sigma_{jj^{\prime}}^{\gamma\gamma^{\prime}(n)}(i\nu_{m})=\sum_{x=a,b,c,...}t_{jj^{\prime}}^{(nx)}\left[J\right]\sigma^{\gamma\gamma^{\prime}(nx)}(i\nu_{m}).\label{eq:sum_x}
\end{equation}

%%%%%%%%%%%%%%%%%%%%%%%%%%%%%%%%%%%%%%%%%%%%%%%%%%%%%%%%%%%%%%%%%%%%%%%%%%
%%%%%%%%%%%%%%%%%%%%%%%%%%%%%%%%%%%%%%%%%%%%%%%%%%%%%%%%%%%%%%%%%%%%%%%%%%
\section{Analytic results for 1-J irreducible Diagrams
\label{sec:Results-for-Sigma}}

In this section we report the analytic results of the diagram evaluation
up to third order in $J$ for the Ising model, TFIM at $T=0$ and
the Heisenberg model as defined in Eqns.~(\ref{eq:Ising}), (\ref{eq:TFIM})
and (\ref{eq:Heisenb}). Diagrams in fourth order and their analytic
results are relegated to App.~\ref{app:DiagramsSigma4}.
%In the diagrams, single-spin blocks representing $\left\langle S_{i}^{\alpha}\right\rangle $ are not required. This is evident in the paramagnetic phase of the Ising and Heisenberg model where these quantities vanish. For the symmetry-unbroken phase of the TFIM they are finite for $\alpha=z$ but these blocks do not appear in the expansion of $G^{xx}=\mathrm{Re}[G^{++}+G^{+-}]$.

We provide results for the second contribution to Eq.~(\ref{eq:Sigma=t*sigma}),
the geometry independent $\sigma^{\gamma\gamma^{\prime}(nx)}(i\nu_{m})$.
For the Ising model and TFIM case we consider $\sigma^{zz}$ and $\sigma^{xx}$,
respectively, see Tab.~\ref{tab:sigma_Ising/TFIM}. In both cases diagrams which
involve blocks of any odd order vanish and are not listed. For the
Ising model this is due to spin flip symmetry $S_{i}^{z}\rightarrow-S_{i}^{z}\;\forall i$
in the paramagnetic phase, for the $\sigma^{xx}$ in the TFIM this
follows from the fact that the blocks solely involve $S^{+}$
and $S^{-}$ which need to appear in equal numbers for any finite
contribution in light of U(1) spin rotation symmetry of $H_{0,h}$.
For the Ising model only static contributions are finite due to its
classical nature. 

The results for $\Sigma_{jj^{\prime}}^{\gamma\gamma^{\prime}(n)}(i\nu_{m})$
computed by summing over topologies {[}Eq.~(\ref{eq:sum_x}){]} have
been tested for small clusters of spins, e.g.~for the dimer with
$N=2$ and $J_{12}=J_{21}=J_{1}$ and $J_{11}=J_{22}=0$. In this
case, for moderately large $S$, the Hilbert space is small and the
exact two-point function $G$ can be found using the spectral representation
\citep{bruusManyBodyQuantum2004}. Hence the exact $\Sigma$ is obtained via Eq.~(\ref{eq:Larkin}). This exact result is then
expanded in $J_{1}$ and checked against $\Sigma_{jj^{\prime}}^{\gamma\gamma^{\prime}(n)}(i\nu_{m})$
computed diagrammatically. As an example, for the TFIM dimer with
$J_{1}>0$, $T=0$ and $S=1/2$, one confirms
\begin{eqnarray}
\Sigma_{11}^{xx}(i\nu) & = & \frac{h}{2\left(h^{2}+\nu^{2}\right)}-J_{1}^{2}\frac{5h^{2}+\nu^{2}}{64h\left(h^{2}+\nu^{2}\right)^{2}}\\
 &  & +J_{1}^{4}\frac{\left(43h^{4}+14h^{2}\nu^{2}+3\nu^{4}\right)}{4096h^{3}\left(h^{2}+\nu^{2}\right)^{3}}+O\left(J_{1}^{6}\right),\nonumber \\
\Sigma_{12}^{xx}(i\nu) & = & J_{1}^{3}\frac{-1}{64h\left(h^{2}+\nu^{2}\right)^{2}}+O\left(J_{1}^{5}\right).
\end{eqnarray}
As some diagrams like (3b) vanish for the dimer geometry it is important to
also check the fully connected trimer with $N=3$ in an analogous fashion. 

Finally, we turn to the Heisenberg case at $T>0$ where the $\sigma^{zz(nx)}=\sigma^{\alpha\alpha(nx)}$
are reported in Tab.~\ref{tab:sigma_Heisenberg}. We need to distinguish
between the static case $\nu_{m}=0$ and the dynamic case $\nu_{m}=2\pi mT\neq0$
for which we abbreviate $\Delta=\frac{1}{2\pi m}$. Again, these results
have been tested for small spin clusters. Another non-trivial check
for the resulting $\Sigma^{(n)}$ is the fulfillment of
$\Sigma_{\mathbf{k}=0}^{(n)}(i\nu_{m}\neq0)=0$ required by the constant-of-motion property of
$S_{\mathbf{k}=0}^{z}$ in a Heisenberg
system.
\begin{table}
\begin{centering}
\begin{tabular}{c|l|l}
 & Ising $T^{1+n}\sigma^{zz(nx)}(i\nu_{m}=0)$ & TFIM $\sigma^{xx(nx)}(i\nu)|_{T=0}$\tabularnewline
\hline 
(0a) & $b_{c,1}$ & $\frac{hS}{h^{2}+\nu^{2}}$\tabularnewline
\hline 
(2a) & $\frac{1}{2}b_{c,1}b_{c,3}$ & $\frac{-S^{2}\left(5h^{2}+\nu^{2}\right)}{16h\left(h^{2}+\nu^{2}\right)^{2}}$\tabularnewline
\hline 
(3b) & $\frac{-1}{2}b_{c,1}^{2}b_{c,3}$ & $\frac{S^{3}\left(4h^{2}+\nu^{2}\right)}{16h^{2}\left(h^{2}+\nu^{2}\right)^{2}}$\tabularnewline
(3c) & $\frac{-1}{6}b_{c,3}^{2}$ & $\frac{-S^{2}}{16\left(h^{2}+\nu^{2}\right)^{2}}$\tabularnewline
\end{tabular}
\par\end{centering}
\caption{\label{tab:sigma_Ising/TFIM}Ising model and TFIM at $T=0$, both
in the symmetric phase: Expansion of the lattice independent
parts $\sigma^{(nx)}$ of the diagrams $\Sigma^{(nx)}$ for $n=0,2,3$ [c.f.~Eq.~(\ref{eq:Sigma=t*sigma})]. Only topologies with finite $\sigma^{(nx)}$ are shown.
The derivatives of the Brillouin function (\ref{eq:bcy}) at zero
field are denoted by $b_{c}^{(m)}(0)\equiv b_{c,m}$. Fourth order
results are given in Tab.~\ref{tab:sigma_4}.}
\end{table}

\begin{table}
\begin{centering}
\begin{tabular}{c|l|l}
$T^{1+n}\sigma^{zz(nx)}$ & static ($\nu_{m}=0$) & dyn. ($\nu_{m}\neq0$)\tabularnewline
\hline 
(0a) & $b_{c,1}$ & $0$\tabularnewline
\hline 
(2a) & $\frac{b_{c,1}^{2}}{-6}\left(1+6b_{c,1}\right)$ & $+2\Delta^{2}b_{c,1}^{2}$\tabularnewline
(2b) & $\frac{b_{c,1}^{2}}{-12}$ & $-2\Delta^{2}b_{c,1}^{2}$\tabularnewline
\hline 
(3a) & $\frac{b_{c,1}^{2}}{-24}\left(1+4b_{c,1}\right)$ & $+\frac{1}{2}\Delta^{2}b_{c,1}^{2}$\tabularnewline
(3b) & $\frac{b_{c,1}^{3}}{6}\left(1+6b_{c,1}\right)$ & $-2\Delta^{2}b_{c,1}^{3}$\tabularnewline
(3c) & $\frac{b_{c,1}^{2}}{-120}\left(48b_{c,1}^{2}+16b_{c,1}+3\right)$ & $-\frac{1}{2}\Delta^{2}b_{c,1}^{2}$\tabularnewline
(3d) & $0$ & $+2\Delta^{2}b_{c,1}^{3}$\tabularnewline
\end{tabular}
\par\end{centering}
\caption{\label{tab:sigma_Heisenberg}Heisenberg model at $T>0$ in the paramagnetic phase: Expansion of the lattice independent parts $\sigma^{(nx)}$ of the
diagrams $\Sigma^{(nx)}$ for $n=0,2,3$, c.f.~Eq.~(\ref{eq:Sigma=t*sigma}).
Fourth order results are given in Tab.~\ref{tab:sigma_4}. The $\Delta$
in the dynamic case stands for $1/(2\pi m)$.}
\end{table}
 
%%%%%%%%%%%%%%%%%%%%%%%%%%%%%%%%%%%%%%%%%%%%%%%%%%%%%%%%%%%%%%%%%
%%%%%%%%%%%%%%%%%%%%%%%%%%%%%%%%%%%%%%%%%%%%%%%%%%%%%%%%%%%%%%%%%
\section{hypercubic lattice benchmarks\label{sec:hypercubic}}

We now proceed to test the applicability of the diagrammatic expansion
for nearest-neighbor models on the hypercubic lattice in $d$ spatial
dimensions with AFM coupling $J_{1}>0$. The geometry factors and further details on the lattice can be found in Tab.~\ref{tab:t023}.
For the Ising and Heisenberg
case we focus on the magnetic ordering temperature $T_{c}$ and for the TFIM at $T=0$ we consider the critical field $h_{c}$ and
the excitation gap $\Delta(h)$. The latter serves as an example for a dynamical
quantity which needs to be evaluated via analytical continuation. We will use the inverse dimension $1/d$ as a small control parameter.

The magnetic phase boundary is signaled by the divergence of the static susceptibility,
according to Eq.~(\ref{eq:Larkin}), 
\begin{equation}
    G_{\mathbf{k}}^{-1}(i\nu_{m}=0)=1/\Sigma_{\mathbf{k}}(i\nu_{m}=0)+J_{\mathbf{k}}\overset{!}{=}0, \label{eq:G=divergent}
\end{equation}
with $\mathbf{k}$ replaced by the N\'eel wavevector $\mathbf{N}=(\pi,\pi,...,\pi)$.
Thus the critical coupling is given by the solution of
$0\overset{!}{=}1/\Sigma_{\mathbf{N}}-2dJ_{1}$.
We specialize to the thermal transition in the Ising case (with $S=1/2$)
or Heisenberg case. First, as stated before, the MF approximation
for $T_{c}$ is found by replacing $\Sigma_{\mathbf{N}}\rightarrow\Sigma_{\mathbf{N}}^{(0)}=\beta b_{c,1}$ in the above equation
which yields the well-known MF result $T_{c}^{(0)}=2 d \, b_{c,1} \, J_{1}\sim d$.
We then divide Eq.~(\ref{eq:G=divergent}) by $T$ and proceed to obtain
corrections to the MF result for $T_{c}$. Following pioneering work in Refs.~\cite{kriegExactRenormalization2019,kriegFunctionalRenormalization2019} we seek a consistent expansion
of 
\begin{equation}
1/(T\Sigma_{\mathbf{N}})=2d\beta J_{1}\label{eq:nnHyTc2}
\end{equation}
in the parameter $1/d$ assumed to be small. In this case MF is approximately valid so that $X_{1}\equiv\beta J_{1}$
close to criticality is of order $\sim1/d$. With this in mind we
provide the hyper-cubic geometry factors $\beta^{n}\tilde{t}_{\mathbf{k}=\mathbf{N}}^{(nx)}$
for all non-vanishing topologies in Tab.~\ref{tab:hypttilde}. Next
to the diagram label we show the leading scaling with $1/d$ of the
particular $\beta^{n}\tilde{t}_{\mathbf{k}=\mathbf{N}}^{(nx)}$ close
to the thermal transition which does depend both on order $n$ and topology $x$.
We can thus expand $1/(T\Sigma_{\mathbf{N}})$ up to order $(1/d)^{m}$,
drop all higher orders and solve numerically for $T_c$. Since
the $\Sigma^{(n)}$ are available up to $n=4$, we can consider $m=0,1,2$,
c.f.~Tab.~\ref{tab:hypttilde}. Explicitly, we have
\begin{align}
\frac{1}{T\Sigma_{\mathbf{N}}} & =\frac{1}{T\sigma^{(0)}}\!-\!\frac{A_{1}}{\left(T\sigma^{(0)}\right)^{2}}\!+\!\frac{A_{2}}{\left(T\sigma^{(0)}\right)^{3}}\!+\!O\left(d^{-3}\right),\label{eq:TcExpansionIn1/d}\\
A_{1} & =2dX_{1}^{2}T^{3}\left(\sigma^{(2a)}-\sigma^{(2b)}\right),\\
A_{2} & =A_{1}^{2}-T^4\sigma^{(0)}[2dX_{1}^{3}\left(\sigma^{(3a)}-\sigma^{(3c)}\right) \\
 & +4d^{2}X_{1}^{4}T(\sigma^{(4b)}-2\sigma^{(4e)}+\sigma^{(4f)}+\sigma^{(4h)}\nonumber \\
 & +3\sigma^{(4j)}+3\sigma^{(4l)}-3\sigma^{(4m)})],\nonumber
\end{align}
where $A_1\sim 1/d$ and $A_2\sim 1/d^2$. 
\begin{table}
\begin{centering}
\begin{tabular}{c|c|c|c|c}
(0a)$\sim\!d^{0}$ & (2a)$\sim\!d^{-1}$ & (2b)$\sim\!d^{-1}$ & (3a)$\sim\!d^{-2}$ & (3c)$\sim\!d^{-2}$\tabularnewline
$1$ & $+2dX_{1}^{2}$ & $-2dX_{1}^{2}$ & $+2dX_{1}^{3}$ & $-2dX_{1}^{3}$\tabularnewline
\hline 
(4a)$\sim\!d^{-3}$ & (4b)$\sim\!d^{-2}$ & (4c)$\sim\!d^{-3}$ & (4e)$\sim\!d^{-2}$ & (4f)$\sim\!d^{-2}$\tabularnewline
$+2dX_{1}^{4}$ & $+4d^{2}X_{1}^{4}$ & $-2dX_{1}^{4}$ & $-8d^{2}X_{1}^{4}$ & $+4d^{2}X_{1}^{4}$\tabularnewline
\hline 
(4h)$\sim\!d^{-2}$ & (4j)$\sim\!d^{-2}$ & (4l)$\sim\!d^{-2}$ & \multicolumn{2}{c}{(4m)$\sim\!d^{-2}$}\tabularnewline
$+4d^{2}X_{1}^{4}$ & $6d\left[2d\!-\!1\right]X_{1}^{4}$ & $6d\left[2d\!-\!1\right]X_{1}^{4}$ & \multicolumn{2}{c}{$-6d\left[2d-1\right]X_{1}^{4}$}\tabularnewline
\end{tabular}
\par\end{centering}
\caption{\label{tab:hypttilde}Non-vanishing geometry factors $\beta^{n}\tilde{t}_{\mathbf{k}=\mathbf{N}}^{(nx)}\left[J\right]$
$(n\protect\leq4$) for the nearest-neighbor $d$-dimensional hyper-cubic
lattice with coupling $J_{1}$ as a function of $X_{1}=\beta J_{1}$.
The leading-order scaling in powers of $1/d$ close to the thermal ordering transition is provided next to
the diagram label. Geometry factors for $n>4$ (not shown) are of order $d^{-3}$ or smaller.}
\end{table}

%%%%%%%%%% ISING
We start with the Ising case, where $T_c$ was already obtained in Ref.~\cite{kriegExactRenormalization2019} up to order $1/d^3$. We report the results for $T_{c}/T_{c}^{(0)}$ and $d=3,...,7$ in Tab.~\ref{tab:IsingHypercubic}. For large $d$, the quasi-exact
results from high-order series expansion or Monte Carlo simulations \cite{ferrenbergPushingLimits2018,lundowCriticalBehavior2009,buteraHightemperatureExpansions2012} are rapidly approached as the order $m$ in $1/d$ is increased.
We also compare our results at order $1/d^{2}$ to other diagrammatic approaches
with different resummation strategies: Our results are similar
to the spin functional RG approach (spin-fRG) by Krieg and Kopietz
\citep{kriegExactRenormalization2019}, only at
$d=3$ the latter has a slight advantage. The dynamical MF theory
for spins \citep{otsukiDynamicalMeanfield2013} (spin-DMFT)
is not competitive. This is not surprising given DMFT's
local approximation of $\Sigma$ already fails in order $J^{3}$ where the non-local diagram (3c) appears \footnote{Note that the lower-order non-local diagram (2b) vanishes for the Ising model}.
\begin{table}
\begin{centering}
\begin{tabular}{c|c|cc|c|c}
$d$ & exact & $O(1/d)$ & $O(1/d^{2})$ & \textcolor{black}{spin-fRG \citep{kriegExactRenormalization2019}} & spin-DMFT\textcolor{blue}{{} \citep{otsukiDynamicalMeanfield2013}}\tabularnewline
\hline 
3 & 0.752 & 0.789 & 0.740 & 0.744 & 0.659\tabularnewline
4 & 0.835 & 0.854 & 0.839 & 0.839 & 0.807\tabularnewline
5 & 0.878 & 0.887 & 0.880 & 0.880 & 0.865\tabularnewline
6 & 0.903 & 0.908 & 0.904 & 0.904 & \tabularnewline
7 & 0.919 & 0.923 & 0.920 & 0.920 & \tabularnewline
\end{tabular}
\par\end{centering}
\caption{\label{tab:IsingHypercubic} Ising model: Critical temperature $T_{c}/T_{c}^{(0)}$
for the $d$-dimensional nearest-neighbor Ising model ($S=1/2$) on
the hypercubic lattice normalized to the MF transition temperature
$T_{c}^{(0)}=d J_{1}/2$. The quasi-exact benchmark
results in the second column \cite{ferrenbergPushingLimits2018,lundowCriticalBehavior2009,buteraHightemperatureExpansions2012} are compared to the results obtained from
Eq.~(\ref{eq:TcExpansionIn1/d}) evaluated up to order $1/d$
and $1/d^{2}$, respectively. For comparison, the last two columns
report results from spin-fRG \citep{kriegExactRenormalization2019}
and spin-DMFT \citep{otsukiDynamicalMeanfield2013}.}
\end{table}

%%%%%%%%%% HEISENBERG
For the Heisenberg model, benchmark checks of $T_{c}$ suffer from
the scarcity of exact results for $d>3$. The exception is the classical case
($S\rightarrow\infty$) for $d=4$. Here $T_{c}/T_{c}^{(0)}$
improves from $0.8536$ in order $1/d$ to $0.8315$ in order $1/d^{2}$
with the exact result at $0.822$ \citep{mckenzieHightemperatureSusceptibility1982}.
For the case $d=3$ and $S<\infty$, Eq.~(\ref{eq:nnHyTc2})
often yields non-real solutions at the limited expansion orders available.
An exception is the case $d=3,\,S=3/2$ for which $T_{c}/T_{c}^{(0)}$
improves from $0.769$ in order $1/d$ to $0.657$ in order $1/d^{2}$
somewhat closer to the exact result $0.702$ \cite{oitmaaCurieNeel2004}.

\begin{figure}
\begin{centering}
\includegraphics{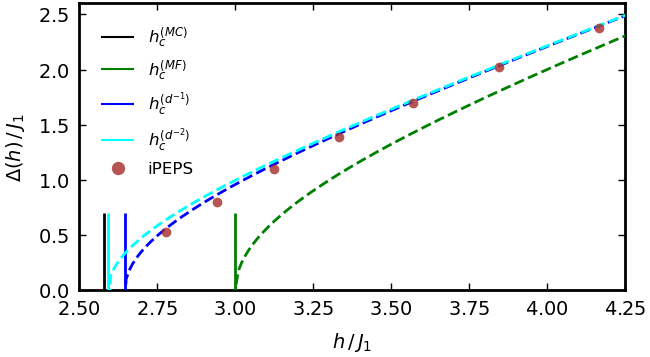}
\par\end{centering}
\caption{\label{fig:TFIM}TFIM on the cubic lattice at $T=0$ and $S=1/2$. The
vertical lines indicate the critical field $h_{c}$ in $1/d$-expansion Eq.~(\ref{eq:nnHyHc}), the quasi-exact result \citep{bloteClusterMonte2002}
(black) is rapidly approached with increasing expansion order.
The same expansion is also used to estimate the spectral gap $\Delta(h)$
shown by dashed lines. The first correction to MF
approximation is already in good agreement to the tensor-network
(iPEPS) simulation of Ref.~\citep{lukinSpectralGaps2024}. 
}
\end{figure}

%%%%%%%%%% TFIM
For the TFIM at $T=0$, the critical
field at the ordering transition is given from Eq.~\eqref{eq:G=divergent} as 
\begin{equation}
1/(d\Sigma_{\mathbf{N}}^{xx})\overset{!}{=}2J_{1}.\label{eq:nnHyHc}
\end{equation}
The MF approximation with $\Sigma_{\mathbf{N}}^{xx}\rightarrow S/h$ yields $h_{c}^{(0)}=2dSJ_{1}\sim d$. Using $\sigma^{xx(nx)}(i\nu=0)|_{T=0} \sim 1/h^{1+n}$ we again expand the left-hand side for $h\gtrsim h_{c}$
in orders of $1/d$ which yields expressions similar to Eq.~(\ref{eq:TcExpansionIn1/d}).
We focus on the case $d=3$, $S=1/2$ and report the results for $h_{c}$
in Fig.~\ref{fig:TFIM}, see vertical lines. The quasi-exact Monte Carlo
result $h_{c}^{(MC)}=2.57907(3)J_{1}$ \citep{bloteClusterMonte2002}
is rapidly approached. 

To showcase the advantage of analytical expressions for the Matsubara
correlator, we consider the excitation gap $\Delta(h)$ for $h\gtrsim h_{c}$
which we obtain from the dynamical $xx$-spin correlator, $G_{\mathbf{k}}^{xx}(i\nu\neq0)$.
The latter contains the dispersion $\omega_{\mathbf{k}}$ of spin waves transversal
to the magnetization in the $S^{z}$ direction determined from the position of the sharp peak in $\mathrm{Im}G_{\mathbf{k}}^{xx,R}(\nu)$. The weight of the peak is the one-particle structure factor. For $h\rightarrow h_{c}$ from above,
we expect $\omega_{\mathbf{k}}$ to vanish at $\mathbf{k}=\mathbf{N}$
where the spin wave softens and order sets in. For $h>h_{c}$, the
gap is thus defined as $\Delta(h)=\omega_{\mathbf{N}}$ and we consider
\begin{equation}
G_{\mathbf{N}}^{xx}(i\nu)=\frac{1/d}{\frac{1}{d\Sigma_{\mathbf{N}}^{xx}(i\nu)}-2J_{1}}.
\end{equation}
We use an expansion of the denominator similar to above which we here
evaluate to order $1/d$ (and analogously for MF and order $1/d^2$). We find
\begin{equation}
G_{\mathbf{N}}^{xx}(i\nu)\!\simeq\!\frac{hS}{h^{2}\!+\!\frac{5}{8}dSJ_{1}^{2}\!-\!2dShJ_{1}\!-\!(i\nu)^{2}\!\! \left(\!1\!+\!\frac{dSJ_{1}^{2}}{8h^{2}}\right)}
\end{equation}
and after analytic continuation the gap is obtained as
\begin{equation}
\Delta^{(1/d)}(h)=\sqrt{{(h^{2}-2dShJ_{1}+\frac{5}{8}dSJ_{1}^{2})}/{(1+\frac{dSJ_{1}^{2}}{8h^{2}}})}.
\end{equation}
Setting $d=3$ and $S=1/2$, this agrees very well with recent iPEPS tensor network
simulations in Ref.~\citep{lukinSpectralGaps2024}, see Fig.~\ref{fig:TFIM} (blue dashed line and dots). For $h \gtrsim h_c$, the iPEPS is actually closer to $\Delta^{(1/d)}(h)$ than to $\Delta^{(1/d^2)}(h)$, we suspect this is an artifact of finite bond dimension of the tensor-network.

%%%%%%%%%%%%%%%%%%%%%%%%%%%%%%%%%%%%%%%%%%%%%%%%%%%%%%%%%%%%%%%%%%%%%%%%%%
%%%%%%%%%%%%%%%%%%%%%%%%%%%%%%%%%%%%%%%%%%%%%%%%%%%%%%%%%%%%%%%%%%%%%%%%%%
\section{Application to models from many-body quantum optics\label{sec:applications}}

%%%%%%%%%%%%%%%%%%%%%%%%%%%%%%%%%%%%%%%%%%%%%%%%%%%%%%%%%%%%%%%%%%%%%%%%%%
\subsection{Long-range square lattice Heisenberg model}
\label{subsec:long-range}

As a first of two applications inspired from many-body quantum optical systems we consider
the $S=1/2$ Heisenberg square lattice model with long-range FM power
law interactions,
\begin{equation}
J_{ii^{\prime}}=J_{1}/|\mathbf{r}_{i}-\mathbf{r}_{i^{\prime}}|^{\alpha},
\label{eq:LongRangeJ}
\end{equation}
with $J_{1}<0$. This model has been recently studied with QMC simulations \citep{zhaoFinitetemperatureCritical2023} which are numerically expensive due to the large system sizes required to approximate the infinite system limit. 
The lattice constant is set to unity and the interesting regime for
the power-law exponent is $\alpha\in(2,4)$ \cite{defenuLongrangeInteracting2023}. In this range $\alpha$
is large enough for a well-defined thermodynamic limit but small enough for a finite FM ordering temperature $T_{c}>0$ evading the Mermin-Wagner theorem \citep{merminAbsenceFerromagnetism1966}. We abbreviate $J_{\mathbf{r}_j}=J_{1}/|\mathbf{r}_{j}|^{\alpha}$.
\begin{figure}
\begin{centering}
\includegraphics{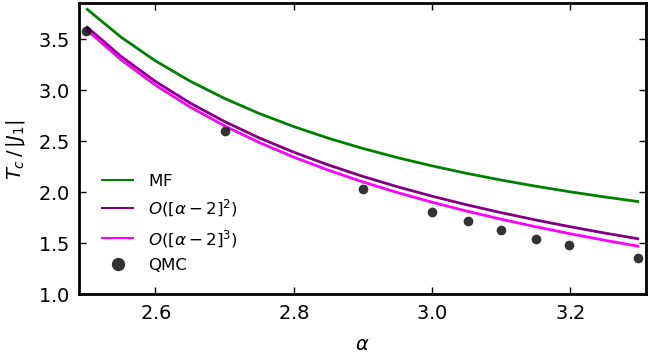}
\par\end{centering}
\caption{\label{fig:LongRange}Ordering temperature for the $S=1/2$ square
lattice Heisenberg FM with couplings decaying as a power law with
exponent $\alpha$, see Eq.~(\ref{eq:LongRangeJ}). Dots denote reference data
from QMC taken from Ref.~\citep{zhaoFinitetemperatureCritical2023}.
Colored lines indicate $T_{c}$ in MF approximation (green), and adding
corrections up to order $[\alpha-2]^{2}$ (purple) and $[\alpha-2]^{3}$ (magenta),
respectively.}
\end{figure}

In analogy to Sec.~\ref{sec:hypercubic}, $T_{c}$ is determined from
\begin{equation}
-\beta J_{\mathbf{k}=\mathbf{0}}\overset{!}{=}1/(T\Sigma_{\mathbf{k}=\mathbf{0}}).\label{eq:TcLongRange}
\end{equation}
To compute the spatial Fourier transform $J_{\mathbf{k}=\mathbf{0}}$
on the left-hand side and for subsequent expansions on the right-hand
side, we obtain numerically the lattice sums $I^{(m)}\equiv\sum_{j\neq0}r_{j}^{-\alpha m}$
for $m=1,2,3$ and a sufficiently large cutoff. Then $J_{\mathbf{k}=\mathbf{0}}=J_{1}I^{(1)}$. To identify a suitable expansion scheme, we consider
the integral approximation of $I^{(m)}$ with lower bound
$a=O(1)$,
\begin{equation}
I^{(m)}\!\simeq\!\!\int_{a}^{\infty}\!\!\!\!\!\mathrm{d}r\,r^{1-\alpha m}\!=\!\frac{a^{2-m\alpha}}{m\alpha\!-\!2}\underset{\alpha\rightarrow2}{\longrightarrow}\!\begin{cases}
\sim\!\frac{1}{\alpha-2}\! & \!\!:m=1,\\
\mathrm{const.} & \!\!:m>1.
\end{cases}
\end{equation}
For $\alpha\rightarrow2$ from above, $I^{(m)}$ only diverges for $m=1$ whereas it is finite for $m>1$. 

From Eq.~\eqref{eq:TcLongRange} the MF critical temperature is
\begin{equation}
T_{c}^{(0)}=|J_{1}|b_{c,1}I^{(1)}\sim\frac{1}{\alpha-2},
\end{equation}
see the green line in Fig.~\ref{fig:LongRange}. We thus use $\alpha-2$ as the small parameter for the expansion on the right-hand side of Eq.~\eqref{eq:TcLongRange}. We consider the geometry factors
$\beta^{n}t_{\mathbf{k}=\mathbf{0}}^{(nx)}$ that play a role for
the static Heisenberg case and we limit ourselves to order $n\leq3$.
For diagram (3b), we numerically checked that $\sum_{i,j}J_{\mathbf{r}_j}J_{\mathbf{r}_i-\mathbf{r}_j}J_{\mathbf{r}_i}\equiv J_{1}^{3}\tilde{I}^{(3)}$
is non-singular for $\alpha\rightarrow2$. In summary, this
means that $T\Sigma^{(n)}\sim \beta^n \sim [\alpha-2]^{n}$ around criticality.
We expand the right-hand side in Eq.~(\ref{eq:TcLongRange})
to third order in $\alpha-2$ and obtain
\begin{align}
 & b_{c,1}\beta|J_{1}|I^{(1)}=1\\
 & -b_{c,1}^{-1}\left(\beta|J_{1}|\right)^{2}I^{(2)}T^3\left(\sigma^{(2a)}+\sigma^{(2b)}\right)\nonumber \\
 & +b_{c,1}^{-1}\left(\beta|J_{1}|\right)^{3}T^4\left[I^{(3)}\left(\sigma^{(3a)}+\sigma^{(3c)}\right)+\tilde{I}^{(3)}\sigma^{(3b)}\right]+...\nonumber 
\end{align}
The results for $S=1/2$ are shown in Fig.~\ref{fig:LongRange} and
approach the QMC data quickly if $\alpha-2$ is sufficiently small.

%%%%%%%%%%%%%%%%%%%%%%%%%%%%%%%%%%%%%%%%%%%%%%%%%%%%%%%%%%%%%%%%5
\subsection{Dicke-Ising model: Ground-state superradiance
\label{subsec:Dicke-Ising-model:-Ground-state}}

As a second application to a many-body quantum optical system we consider the
Dicke-Ising model \citep{zhangQuantumPhases2014,gelhausenQuantumopticalMagnets2016,rohnIsingModel2020,schellenbergerAlmostEverything2024a}. This also provides an example where single-spin blocks appear in the diagrammatic expansion. This means that the pre-computed diagrams of Sec.~\ref{sec:Results-for-Sigma} cannot be used in the following calculation, see below.

\begin{figure}
\begin{centering}
\includegraphics{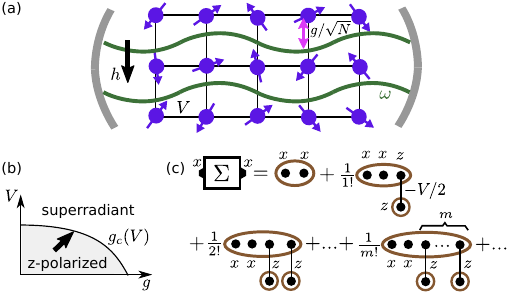}
\par\end{centering}
\caption{\label{fig:Dicke-Ising}Dicke-Ising model: (a) Sketch of an experimental
square lattice setup with cavity photons at frequency $\omega$ coupled to spins with strength $g/\sqrt{N}$, homogeneous $z$-field $h$ and nearest-neighbor AFM Ising interactions $V$ induced via Rydberg dressing. (b) Schematic
phase diagram in the vicinity of the $z$-polarized phase and
the adjacent superradiant phase in the presence of AFM Ising interactions
$V$. (c) The diagrammatic expansion for $\Sigma^{xx}$ in the limit
$N\rightarrow\infty$ and $T\rightarrow0$ can be summed exactly.
This shows that the exact phase boundary in (b) is described by the MF
result Eq.~\eqref{eq:gc}. }
\end{figure}

The Hamiltonian features a competition between a homogeneous field $h>0$ in $z$-direction, a coupling of the total spin-$x$ to a cavity photon and
a nearest-neighbor AFM Ising $zz$-interaction $V$, see Fig.~\ref{fig:Dicke-Ising}(a) for a sketch. For concreteness, we specialize to a square lattice geometry. However, our qualitative results below do not depend on this choice.
The Hamiltonian is $H=H_{x}+H_{z}$ with
\begin{eqnarray}
H_{x} & = & \omega a^{\dagger}a+\frac{g}{\sqrt{N}}\left(a+a^{\dagger}\right)\sum_{i}S_{i}^{x},\\
H_{z} & = & -h\sum_{i}S_{l}^{z}+\frac{V}{2}\sum_{\left\langle i,i^{\prime}\right\rangle }\!\left(S_{i}^{z}+\frac{1}{2}\right)\!\left(S_{i^{\prime}}^{z}+\frac{1}{2}\right) \label{eq:Hz1}\\
 & = & -\left[h-V\right]\sum_{i}S_{i}^{z}+\frac{V}{2}\sum_{\left\langle i,i^{\prime}\right\rangle }S_{i}^{z}S_{i^{\prime}}^{z}. \label{eq:Hz2}
\end{eqnarray}
Here, the cavity photon at frequency $\omega$ is created by $a^\dagger$. The sums go over $N$
sites and the nearest neighbor bonds, respectively. The cavity-dipole coupling is $g/\sqrt{N}>0$. This ensures an
extensive interacting energy in the infinite system limit $N\rightarrow\infty$
which we consider in the following. 

The Dicke model (the case $V=0$) and its phase transition
to the symmetry-broken superradiant state (the FM state with $\left\langle S_{i}^{x}\right\rangle \neq0$)
has been studied thoroughly both in theory \citep{heppSuperradiantPhase1973,kirtonIntroductionDicke2019}
and experiment \citep{baumannDickeQuantum2010}, the latter in a non-equilibrium
setting. However, comparatively little is known about the
Dicke-Ising model ($V>0$) \cite{zhangQuantumPhases2014,gelhausenQuantumopticalMagnets2016,rohnIsingModel2020,schellenbergerAlmostEverything2024a} which is yet awaiting experimental implementation. 
A possible realization for the Ising $zz$-interactions uses the concept of Rydberg dressing \citep{pupilloRydbergDressed2010} which motivated the form of Eq.~\eqref{eq:Hz1} and leads to a renormalized effective field $h\rightarrow h-V$ when rewritten as in Eq.~\eqref{eq:Hz2}. 

So far, the model also contains bosonic degrees of freedom in variance to the spin-only formulation in Hamiltonian~\eqref{eq:Haniso_h}. However, assuming a thermal state we can trace out the photons on the level
of the generating functional \citep{weberQuantumMonte2022}. This
replaces $H_{x}$ by an all-to-all retarded (i.e.~frequency
dependent) FM spin interaction of the $xx$-type,
\begin{equation}
H_{x}\!\rightarrow \! H_{x,ret}=T\!\sum_{\nu_{m}}\frac{1}{2}\sum_{i,i^{\prime}}S_{i}^{x}(-\nu_{m})J_{ii^{\prime}}^{xx}(\nu_{m})S_{i^{\prime}}^{x}(\nu_{m}), \label{eq:ret-xx}
\end{equation}
with $J_{ii^{\prime}}^{xx}(\nu_{m})=-2g^{2}\omega/[N(\nu_{m}^{2}+\omega^{2})].$
This is close to Eq.~\eqref{eq:Haniso_h}, only differing in the frequency dependence and the presence of an onsite term. As we will discuss, this is inconsequential in the following since no $xx$-interactions will appear in the 1JI diagrammatic expansion.

Here, we focus
on thermal equilibrium at $T\rightarrow0$ and limit the investigation
to the realistic case of small $V$ ($V\ll h$) and its effect on the 
transition between the $z$-polarized and superradiant phase. For $V\rightarrow 0$, it is well known that the MF approximation is exact for the Dicke model \cite{kirtonIntroductionDicke2019} but what happens at $V>0$? As a first step, in Ref.~\citep{gelhausenQuantumopticalMagnets2016} the phase boundary for the Dicke-Ising model on the square lattice was determined in MF approximation,
\begin{equation}
g_{c}=\sqrt{\omega\left(h-2V\right)},\label{eq:gc}
\end{equation}
see Fig.~\ref{fig:Dicke-Ising}(b) for a sketch.

Two questions are in order: First, is the phase transition indeed continuous as in MF approximation? And second, if yes, what are the corrections to the phase boundary beyond the MF prediction? 

As shown in Ref.~\cite{rohnIsingModel2020} for a simplified and more symmetric version of the Dicke-Ising model with $h-V=0$ (with no $z$-polarized phase), the transition between the superradiant and Ising phase is of first order in a chain geometry. However, it is currently unknown if and under which conditions this also holds for the case $V\ll h$ and the phase transition out of the fully polarized phase that we consider. 

Making the assumption that the phase transition from the $z$-polarized to the superradiant phase is continuous (and can thus be detected via a divergence in spin-susceptibility) we use the spin diagrammatic technique to show rigorously our
main finding of this section: Eq.~(\ref{eq:gc}) is already the exact
result. This conclusion has been obtained independently in the very recent work of Schellenberger and Schmidt \cite{schellenbergerAlmostEverything2024a} using an alternative algebraic approach. 

Analogous to the other examples in this work, we start from the $z$-polarized symmetry-unbroken phase and approach the boundary of the superradiant phase, 
see the arrow in Fig.~\ref{fig:Dicke-Ising}(b). We detect the critical
light-matter coupling strength $g_{c}$ from the divergence of the static FM correlator
$G_{\mathbf{k}=\mathbf{0}}^{xx}(i\nu=0)$,
\begin{equation}
1/\Sigma_{\mathbf{k}=\mathbf{0}}^{xx}+J_{\mathbf{k}=\mathbf{0}}^{xx}\overset{!}{=}0\label{eq:Dicke-Ising_gc}
\end{equation}
Here and in the following we drop the zero-Matsubara frequency argument
from all quantities. The static part of the retarded $xx$-interaction \eqref{eq:ret-xx}
is $J_{\mathbf{k}=\mathbf{0}}^{xx}=-2g^{2}/(\omega N)$ and $\Sigma^{xx}$
denotes the $J^{xx}$-irreducible part of $G^{xx}$. Next, we expand
$\Sigma^{xx}$ in $J^{xx}$ and $V$ with the non-interacting Hamiltonian
being $H_{0,h-V}=-\left[h-V\right]\sum_{i}S_{i}^{z}$. The first observation
is that due to the limit $N\rightarrow\infty$ any occurrence of an
interaction $J_{ii^{\prime}}^{xx}\sim1/N$ needs to be accompanied
with a free site summation. Since these diagrams are necessarily $J^{xx}$-reducible
(c.f.~the discussion at the end of Sec.~\ref{subsec:Diagrams,1JI,MF})
they do not occur in $\Sigma^{xx}$ and only the $V$-interactions
$\sim S_{i}^{z}S_{i^{\prime}}^{z}$ need to be considered. The second
observation pertains to $z$-only blocks $G_{0,c,h-V}^{z...z}\left(\tau_{1},...,\tau_{m}\right)=b_{c}^{(m-1)}(\beta[h-V])$.
In the limit $T\rightarrow0$ where the spins are fully polarized,
these generalized susceptibilities vanish except for $m=1$ for which $G_{0,c,h-V}^{z}\left(\tau_{1}\right)=S=1/2$. 

These considerations equate the exact (and local) $\Sigma^{xx}$
to the infinite sum of diagrams shown in Fig.~\ref{fig:Dicke-Ising}(c).
The calculation proceeds without the kernel trick since all diagrammatic
objects are free of frequency loops and are easily evaluated at the required zero frequency.
We only need $\Sigma^{xx(0)}=1/(2[h-V])$, see Tab.~\ref{tab:sigma_Ising/TFIM}
for the TFIM. The fully static mixed-flavor blocks with two $S^{x}$
operators and $m$ appearances of the $S^{z}$ operator can be found
via a $m$-fold derivative of $G_{0,c,h-V}^{xx}$ with respect to $h$ (which thus is also used as a source field),
\begin{equation}
G_{0,c,h-V}^{xx\overset{m}{\overbrace{z...z}}}=\partial_{h}^{m}G_{0,c,h-V}^{xx}=\frac{(-1)^{m}m!}{2[h-V]^{m+1}}.
\end{equation}
With these preparations the infinite diagrammatic sum in Fig.~\ref{fig:Dicke-Ising}(c)
results in the \emph{exact} expression
\begin{eqnarray}
\Sigma_{\mathbf{k}=\mathbf{0}}^{xx} & = & \sum_{m=0}^{\infty}\frac{1}{m!}G_{0,c,h-V}^{xx\overset{m}{\overbrace{z...z}}}\cdot\left\langle S^{z}\right\rangle _{0,c}^{m}\cdot\left(-4\frac{V}{2}\right)^{m}\\
 & = & \frac{1}{2(h-2V)}.\nonumber 
\end{eqnarray}
Inserting this in Eq.~(\ref{eq:Dicke-Ising_gc}) we obtain that the
exact $g_{c}$ is already given by Eq.~(\ref{eq:gc}).

%Although the exactness of the MF approximation for the situation considered is a new result which nicely demonstrates the versatility of the spin diagrammatic approach, our finding can be rationalized in a simple alternative way. First, the MF nature of the cavity induced interaction should not be surprising due to its all-to-all nature, a formal mathematical proof has been given long ago in Ref.~\citep{denoudenSystemsSeparable1976}. Using this, $H=H_{x}+H_{z}$ turns into a TFIM with longitudinal field. The transverse field is only switched on in the superradiant phase {[}c.f.~Fig.~\ref{fig:Dicke-Ising}(b){]} so that in the z-polarized phase a longitudinal field Ising model remains. Finally, as Ising models have product ground states, a MF decoupling of the zz-interaction is necessarily exact for $T\rightarrow0$. 

For future work on the Dicke-Ising model, an extension to the complete
phase diagram which includes also a $z$-AFM and a combined $z$-AFM and
superradiant $x$-FM phase, would be interesting. According to Ref.~\cite{schellenbergerAlmostEverything2024a}, MF theory is again exact for the transition between the latter two phases. 
Likewise, we suggest to consider
the experimentally relevant modifications to finite $N$ and the
open-system case \citep{torreKeldyshApproach2013}. 

%%%%%%%%%%%%%%%%%%%%%%%%%%%%%%%%%%%%%%%%%%%%%%%%%%%%%%%%%%%%%%%%%%%%%
%%%%%%%%%%%%%%%%%%%%%%%%%%%%%%%%%%%%%%%%%%%%%%%%%%%%%%%%%%%%%%%%%%%%%
\section{Conclusion\label{sec:conclusion}}

In summary, we have presented an analytic approach
to Matsubara spin-spin correlation functions based on a diagrammatic expansion
of their 1-J-irreducible part $\Sigma$ to $n$-th order in $J$. We provide
closed-form expressions for $n\leq 4$ for Ising, TFIM and Heisenberg models of completely general lattice geometry and spin length $S$. The introduction of the kernel
function trick was instrumental in this calculation. The final
results are conveniently tabulated for forthcoming application in diverse
contexts where other computationally much more involved methods like tensor networks or QMC are at their limits.

Via many examples and by applying a composite expansion strategy involving the inverse spatial dimension (or similar) as a small parameter, we showed the quantitative success of the diagrammatic approach if applied
to models qualitatively described by the MF approximation. We argued
that this is often the case in highly-connected spin models relevant
for state-of-the-art many-body quantum optical experiments. We use a long-range
Heisenberg model and the Dicke-Ising model, both on the square lattice,
as a showcase. Moreover we provided various benchmark examples for
nearest-neighbor models on the (hyper-)cubic lattice where our method
yields accurate magnetic phase boundaries (both at $T>0$ for Ising
and Heisenberg models and $T=0$ for TFIM). We emphasize that
due to the analytic nature of our approach, continuation to real frequencies is
easily performed. For example we showed competitive results for
the gap in the TFIM. 

Future work could extend our approach to yet higher orders in $J$ (for which diagram creation can be automated) or a greater variety of spin models, including those with non-trivial unit cells or various forms of disorder. Also spin-spin couplings between sites $i,i^\prime$ that are characterized by two or more non-zero parameters in the $3\times 3$ coupling matrix $J_{ii^\prime}^{\gamma\gamma^\prime}$ [c.f.~Eq.~\eqref{eq:Haniso_h}] like e.g.~XXZ models \cite{peterAnomalousBehavior2012} could be studied with little extra effort.
Another option would be to extend our approach to systems with $\mathrm{SU}(N)$ symmetry for $N>2$ \citep{cazalillaUltracoldFermi2014,zhangSpectroscopicObservation2014}. Also, the treatment of the symmetry-broken phase for the study of magnetization and spin-wave properties
\citep{gollSpinFunctional2019,izyumovLongitudinalSpin2002} is within reach.
Further, it would be interesting to consider analytic continuation beyond the computation of the gap to obtain spectral functions. These are routinely measured in inelastic neutron scattering on
solid-state magnets \citep{chenPhaseDiagram2024} or, more recently via
quench spectroscopy in Rydberg tweezer arrays \citep{chenSpectroscopyElementary2023}.
For the static case, analytic insights offered by our approach
proved essential to shed new light \citep{schneiderTamingSpin2024}
onto the puzzle of quantum-to-classical correspondence for static
spin correlation functions in $d>1$ dimensions \citep{kulaginBoldDiagrammatic2013}.

Finally, we point out that the combination of the spin-spin correlator's bare series expansion with the kernel function trick [c.f.~Eq.~\eqref{eq:G^(n)KernelS}] is suitable for evaluation by a diagrammatic Monte Carlo approach \cite{burkardManuscriptPreparation} similar to the connected determinant method \cite{rossiDeterminantDiagrammatic2017}. However imaginary time integrals are treated exactly via the kernel functions. 
Implementing these ideas would enhance the available expansion orders in spin diagrammatics and allow for flexible resummation
schemes beyond this work.

\emph{Note added:} During the completion of the manuscript, we became
aware of independent work in Ref.~\citep{ruckriegelRecursiveAlgorithm2024}
which computes $\Sigma^{(0,2,3)}$ for the Heisenberg model in the
traditional diagrammatic way via frequency integrals. The diagrams were derived from an expansion of spin-fRG flow equations and the five-point free spin correlator was provided analytically.
Applications concern the chiral non-linear susceptibility and estimates of $T_{c}$ for the $d=3,4$ hyper-cubic
case for various $S$. However, due to the chosen expansion of $1/\Sigma$ in $J$,
the quality of the results for $T_c$ does not improve with expansion order,
in contrast to the composite expansion strategies presented in this
work. 

%%%%%%%%%%%%%%%%%%%%%%%%%%%%%%%%%%%%%%%%%%%%%%%%%%%%%%%%%%%%%%%%%%%%
%%%%%%%%%%%%%%%%%%%%%%%%%%%%%%%%%%%%%%%%%%%%%%%%%%%%%%%%%%%%%%%%%%%%
\section{Acknowledgment}

We thank Marin Bukov, Elio König, Peter Kopietz, Andreas Rückriegel,
Achim Rosch, Johannes Reuther, Kai Schmidt, Nils Schopohl and Sebastian Slama for useful discussions
and the authors of Ref.~\citep{lukinSpectralGaps2024} for sharing
their iPEPS data. 
We acknowledge funding from the Deutsche Forschungsgemeinschaft (DFG, German Research Foundation) through the Research Unit FOR 5413/1, Grant No. 465199066. 
B.Sch.~acknowledges funding from the Munich Quantum Valley, supported by the
Bavarian state government with funds from the Hightech Agenda Bayern Plus.
B.Sb.~and B.Sch.~are supported by DFG grant no.~524270816. 
I.L. acknowledges financing from the Baden-W\"urttemberg Stiftung through Project No.~BWST\_ISF2019-23

\appendix
\onecolumngrid

%%%%%%%%%%%%%%%%%%%%%%%%%%%%%%%%%%%%%%%%%%%%%%%%%%%%%%%%%%%%%%%%%%%%%%
\section{V-connected correlators}
\label{app:V-con}

We define the $V$-connected correlators \cite{rossiDeterminantDiagrammatic2017} (subscript $V\!-\!c$) that appear in the formal expansion of the spin correlator in Eq.~\eqref{eq:G_series} and the following equations. For brevity, we set $S_{j}^{\gamma}(\tau)S_{j^{\prime}}^{\gamma^{\prime}}(\tau^{\prime})=A$. Then the $V$-connected correlators
are defined recursively via
\begin{equation}
\left\langle \mathcal{T}V(\tau_{1})...V(\tau_{n})A\right\rangle _{0,h,V \mkern-4mu -\mkern-1.5mu c} 
=
\left\langle \mathcal{T}V(\tau_{1})...V(\tau_{n})A\right\rangle _{0,h} \! -
\sum_{S\subsetneq\{1,...,n\}}
\left\langle \mathcal{T}   \prod_{j\in S}V(\tau_{j})
A\right\rangle _{0,h,V \mkern-4mu -\mkern-1.5mu c} 
\left\langle \mathcal{T} \prod_{k\in\{1,..,n\}\backslash S}V(\tau_{k}) \right\rangle _{0,h}.\label{eq:<...>V-con}
\end{equation}
Note that the ordering of (bosonic) operators behind imaginary time-ordering operator $\mathcal{T}$ does not matter. Standard connected spin correlators [with subscript $c$ first appearing in Eq.~\eqref{eq:G0con}] are defined in analogy to
their $V$-connected counterparts in Eq.~\eqref{eq:<...>V-con} by replacing each $V(\tau_j)$ with a single spin operator and removing the external operators $A \rightarrow 1$.

%%%%%%%%%%%%%%%%%%%%%%%%%%%%%%%%%%%%%%%%%%%%%%%%%%%%%%%%%%%%%%%%%
\section{Evaluation of $\sigma^{\gamma \gamma^\prime(nx)}(i\nu_m)$ via kernel function trick}
\label{app:KernelFunction}

To facilitate all calculations on the right-hand side of Eq.~(\ref{eq:G(nu_m)_series}) and in particular the evaluation of $\sigma^{(nx)}$ we introduce what we call the kernel function trick.
Originally, kernel functions have been introduced to link the Fourier
transform of an imaginary time-ordered $m$-point correlation function
$G_{A_{1}...A_{m}}\left(i\omega_{1},...,i\omega_{m-1}\right)$ to eigenstates and -energies $H\left|\underline{a}\right\rangle =E_{\underline{a}}\left|\underline{a}\right\rangle $
of the many-body Hamiltonian \citep{kuglerMultipointCorrelation2021}. The $m$-point correlators are a generalization of Eq.~(\ref{eq:G(=Cnu_m)}) to $m$ arbitrary
operators $A_{1,2,...,m}$. For the frequency arguments we introduced an abbreviated notation where $\omega_{1}$ is short
for $\omega_{n_{1}}$ and so on. 
The kernel functions straightforwardly extend the well-known spectral (or Lehmann)
representation of the $2$-point correlator \citep{bruusManyBodyQuantum2004} to the $m$-point case.
For bosonic (and spin) operators, the Fourier transform reads
\citep{halbingerSpectralRepresentation2023}
\begin{equation}
G_{A_{1}...A_{m}}\left(i\omega_{1},...,i\omega_{m-1}\right)=\frac{1}{Z}\sum_{p\in S_{m}}\sum_{\underline{1}...\underline{m}}e^{-\beta E_{\underline{1}}}A_{p(1)}^{\underline{1}\underline{2}}A_{p(2)}^{\underline{2}\underline{3}}...A_{p(m)}^{\underline{m}\underline{1}}K_{m}\left(\Omega_{p(1)}^{\underline{1}\underline{2}},\Omega_{p(2)}^{\underline{2}\underline{3}},...,\Omega_{p(m-1)}^{\underline{m-1}\underline{m}}\right),
\label{eq:Gm_Kernel}
\end{equation}
where $A_{k}^{\underline{a}\underline{b}}=\left\langle \underline{a}|A_{k}|\underline{b}\right\rangle $
are matrix elements and the argument of the kernel function $K_{m}$
is a list of $m$ complex numbers $\Omega_{k}^{\underline{a}\underline{b}}\equiv i\omega_{k}+E_{\underline{a}}-E_{\underline{b}}$
which sum to zero so that the last one is often dropped as in Eq.~(\ref{eq:Gm_Kernel}).
The kernel functions itself are completely universal and do neither
depend on the Hamiltonian $H$ nor on the operators $A_{k}$ in Eq.~(\ref{eq:Gm_Kernel}). For example, $K_2(\Omega_1)=-\Delta_{\Omega_1}+\beta\delta_{\Omega_1}/2$ where $\delta_x=\delta_{0,x}$ and $\Delta_x=(1-\delta_x)/x$. 
The recent advance in Ref.~\citep{halbingerSpectralRepresentation2023}
was the calculation of $K_{m}$ for general $m$.

In the context of diagram evaluation, the crucial insight is that Eq.~\eqref{eq:Gm_Kernel} and $n$-th order
perturbative expressions are naturally
connected by interpreting the right-hand side of Eq.~(\ref{eq:G(nu_m)_series})
as a Fourier transform of a time-ordered correlator of order
$n+2$ with the first $n$ frequencies being zero. Hence, as the main technical result of this work,
Eq.~(\ref{eq:G(nu_m)_series}) is expressed as
\begin{eqnarray}
G_{jj^{\prime}}^{\gamma\gamma^{\prime}(n)}\!(i\nu_{m}) &= &\frac{(-1)^{n}}{n!}\sum_{p\in S_{n+2}} \;
\sum_{i_{1}<i_{1}^{\prime},...,i_{n}<i_{n}^{\prime}} \;
\sum_{\gamma_{1,...,n}^{(\prime)}\in\{+,-,z\}} \;
J_{i_{1}i_{1}^{\prime}}^{\gamma_{1}\gamma_{1}^{\prime}}...J_{i_{n}i_{n}^{\prime}}^{\gamma_{n}\gamma_{n}^{\prime}}\label{eq:G^(n)KernelS}\\
&\times &\left\langle B_{p(1)}B_{p(2)}...B_{p(n+2)}\right\rangle _{0,h,V \mkern-4mu -\mkern-1.5mu c}\!K_{n+2}\!\left(\Omega_{p(1)},\Omega_{p(2)},...,\Omega_{p(n+1)}\right).\nonumber
\end{eqnarray}
Here we replaced $V$ using Eq.~(\ref{eq:H0h_V}) and defined the following operator and complex-frequency
lists
\begin{eqnarray}
\left\{ B_{1},...,B_{n},B_{n+1},B_{n+2}\right\}  & = & \left\{ S_{i_{1}}^{\gamma_{1}}S_{i_{1}^{\prime}}^{\gamma_{1}^{\prime}},...,S_{i_{n}}^{\gamma_{n}}S_{i_{n}^{\prime}}^{\gamma_{n}^{\prime}},S_{j}^{\gamma},S_{j^{\prime}}^{\gamma^{\prime}}\right\} ,\label{eq:B_list} \\
\left\{ \Omega_{1},...,\Omega_{n},\Omega_{n+1},\Omega_{n+2}\right\}  & = & \left\{ -(\gamma_{1}+\gamma_{1}^{\prime})h,...,-(\gamma_{n}+\gamma_{n}^{\prime})h,i\nu_{m}-\gamma h,-i\nu_{m}-\gamma^{\prime}h\right\} .\label{eq:=COmega_list_app}
\end{eqnarray}
Note that the external indices on the left-hand side of Eq.~\eqref{eq:G^(n)KernelS} determine the last two entries of the lists. The $\Omega$-list \eqref{eq:=COmega_list_app} was already given in Eq.~\eqref{eq:=COmega_list}. It is to be understood with
the following replacement rule of flavor labels by numbers: $\{z,+,-\}\rightarrow\{0,+1,-1\}$.
The simple structure of $H_{0,h}$ with its
many-body (product) eigenstates and ladder-like energies 
is essential in the derivation of Eq.~\eqref{eq:G^(n)KernelS} as it allows to reduce the complexity of the general Eq.~(\ref{eq:Gm_Kernel})
by rewriting the sum over eigenstates via the equal-time free spin correlator. The point is that the $\Omega_{k}^{\underline{a}\underline{b}}\equiv i\omega_{k}+E_{\underline{a}}-E_{\underline{b}}$
only depend on the flavor(s) $\{+,-,z\}$ of the (composite) operator
$B_{k}$.

Further, we rewrite Eq.~(\ref{eq:G^(n)KernelS}) using an index
ordering operator $\mathcal{P}$. This operator applies to operator
strings and argument lists alike. It acts like time-ordering, but
for discrete indices $(1),(2),...,(n+2)$ that - unlike imaginary
time arguments - do not affect the operator, for example $\mathcal{P}B_{1}(1)B_{2}(3)B_{3}(2)=B_{2}B_{3}B_{1}$.
We also reinstate the redundant last argument $\Omega_{p(n+2)}$ of
$K_{n+2}$. We obtain
\begin{align}
G_{jj^{\prime}}^{\gamma\gamma^{\prime}(n)}(i\nu_{m}) & =\frac{(-1)^{n}}{n!}\sum_{p\in S_{n+2}} \; \sum_{i_{1}<i_{1}^{\prime},...,i_{n}<i_{n}^{\prime}}
\; \sum_{\gamma_{1,...,n}^{(\prime)}\in\{+,-,z\}}\!\!\!\!\!J_{i_{1}i_{1}^{\prime}}^{\gamma_{1}\gamma_{1}^{\prime}}...J_{i_{n}i_{n}^{\prime}}^{\gamma_{n}\gamma_{n}^{\prime}} \label{eq:GwithP}\\
 & \times
 \left\langle \mathcal{P}B_{1}(p_{1})B_{2}(p_{2})...B_{n+2}(p_{n+2})\right\rangle _{0,h,V \mkern-4mu -\mkern-1.5mu c}K_{n+2}\left(\mathcal{P}\left\{ \Omega_{1}(p_{1}),\Omega_{2}(p_{2}),...,\Omega_{n+2}(p_{n+2})\right\} \right),\nonumber 
\end{align}
As the operators $B_{1},B_{2},...,B_{n+2}$ and their associated $\Omega_{1},\Omega_{2},...,\Omega_{n+2}$
appear in all possible orderings, this is evidently the same as Eq.~(\ref{eq:G^(n)KernelS}). 

To calculate $\sigma^{\gamma \gamma^\prime(nx)}(i\nu_m)$ for a particular 1JI diagram $(nx)$ defined by reference site-configuration
$\{i_{k}^{(nx)}<i_{k}^{\prime(nx)}\}_{k=1,2,..,n}$, we specialize
the site-sums in Eq.~\eqref{eq:GwithP} to this reference configuration and split off the geometry factor $t^{(nx)}[J]$ as explained in Sec.~\ref{subsec:diagEval}.
The V-connected average then becomes an ordinary connected equal-time
average with respect to $H_{0,h}$ which factorizes according to the
blocks of equal sites characteristic for $(nx)$. Examples for diagrams $(2a)$ and $(3b)$ are provided in Eqns.~\eqref{eq:(2a)} and \eqref{eq:(3b)}, respectively.

%%%%%%%%%%%%%%%%%%%%%%%%%%%%%%%%%%%%%%%%%%%%%%%%%%%%%%%%%%%%%%%%%%%%%%
\section{Simplified kernel functions for the special cases $h=0$ and $T=0$}
\label{app:Special-Kernel-functions}

Kernel functions $K_{k}\left(\Omega_{1},\Omega_{2},...,\Omega_{k}\right)$
for general complex arguments (obeying $\Omega_{1}+...+\Omega_{k}=0$)
and arbitrary $k=2,3,4,5,6...$ are provided in Ref.~\citep{halbingerSpectralRepresentation2023}.
However, in the context of this paper, where kernel functions are
applied in the framework of spin perturbation theory, we have only two
potentially non-real entries in the $\Omega$-list \eqref{eq:=COmega_list_app}
(the ones at the end which contain $\pm i\nu_{m}$ related to the external operators).
More importantly, we limit ourselves to two special cases, (i) the case $h=0$ for the
Ising and Heisenberg model and (ii) the
limit $T\rightarrow0$ for the TFIM. In these cases substantial simplifications
arise.

(i) Case $h=0$: In Eq.~\eqref{eq:GwithP}, the arguments of $K_k$
are zero except for a possible non-zero pair of frequencies $\pm i\nu_{m}=\pm2m\pi Ti$
shuffled to positions $a_{1,2}$ in the list of length $k=n+2$,
\begin{equation}
\left\{ \Omega_{1},\Omega_{2},...,\Omega_{k}\right\} =(0,0,...,0,\underset{\mathrm{pos.}a_{1}}{\underbrace{i\nu_{m}}},0,...,0,\underset{\mathrm{pos.}a_{2}}{\underbrace{-i\nu_{m}}},\underset{k-a_{2}}{\underbrace{0,...,0}}),
\end{equation}
where we assume $a_{1}<a_{2}$ without loss of generality (see below).
For this situation, we define
\begin{equation}
K_{k}\left(\Omega_{1},\Omega_{2},...,\Omega_{k}\right)\equiv K_{k}^{(h=0)}\left(a_{1},a_{2},m\right).
\end{equation}
From Ref.~\citep{halbingerSpectralRepresentation2023}, we find after
some algebra
\begin{equation}
T^{k-1}K_{k}^{(h=0)}\left(a_{1},a_{2},m\right)=\begin{cases}
\frac{1}{k!} & :m=0,\\
(-1)^{a_{2}-a_{1}}\sum_{l=a_{2}-a_{1}}^{k-a_{1}}\frac{\left[\Delta_{2\pi mi}\right]^{l}}{(k-l)!}\left(\begin{array}{c}
l-1\\
a_{2}-a_{1}-1
\end{array}\right) & :\mathrm{otherwise},
\end{cases}\label{eq:Kernel(h=00003D0)}
\end{equation}
where $\Delta_{x}=1/x$ for non-zero $x$ and zero otherwise. For
the case that $-i\nu_{m}$ appears first, $a_{1}>a_{2}$, we can flip
the sign of $m$ and obtain $K_{k}\left(\Omega_{1},\Omega_{2},...,\Omega_{k}\right)=K_{k}^{(h=0)}\left(a_{2},a_{1},-m\right)$.

(ii) Case $T\rightarrow0$: Here the kernel functions simplify because
certain sums are dominated by inverse temperature $\beta\rightarrow\infty$.
Any $\beta\delta_{\omega_{1}+...+\omega_{l},0}$ that remains must be interpreted as
$\beta\delta_{\omega_{1}+...+\omega_{l},0}\overset{T\rightarrow0}{\longrightarrow}2\pi\delta\left(\omega_{1}+...+\omega_{l}\right)$,
but this does not appear for the particular correlators we compute
in this work. To express the resulting $K_{k}^{(T\rightarrow0)}\left(\Omega_{1},...,\Omega_{k}\right)$
we define the list of the partial sums 
\begin{equation}
\{\Omega_{1},\Omega_{1}+\Omega_{2},\Omega_{1}+\Omega_{2}+\Omega_{3},...,\Omega_{1}+...+\Omega_{k-1}\}\equiv\{c_{1},c_{2},...,c_{k-1}\}.\label{eq:clist}
\end{equation}
The final expression for $K_{k}^{(T\rightarrow0)}$ involves a product of all but the $l$ entries of \eqref{eq:clist} which are zero,
\begin{equation}
K_{k}^{(T\rightarrow0)}\left(\Omega_{1},...,\Omega_{k}\right)=(-1)^{k+1+l}\frac{\beta^{l}}{(l+1)!}\prod_{c_{m}\neq0}\frac{1}{c_{m}}.\label{eq:Kernel(T=00003D0)}
\end{equation}

%%%%%%%%%%%%%%%%%%%%%%%%%%%%%%%%%%%%%%%%%%%%%%%%%%%%%%%%%%%%%%%%%%%%%5
\section{Diagrams and results for $\Sigma^{(4)}$}
\label{app:DiagramsSigma4}

\begin{figure*}[h]
\begin{centering}
\includegraphics{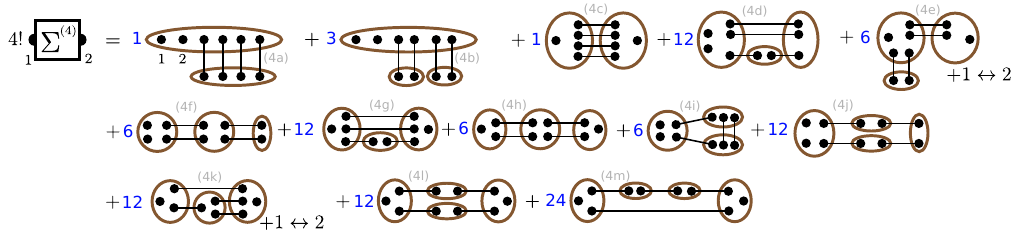}
\par\end{centering}
\caption{\label{fig:=CSigma(4)}Diagrams for $\Sigma$ of order $J^{4}$,
analogous to Fig.~\ref{fig:Sigma023-Heisenberg}. The
label $+\!1\!\leftrightarrow\!2$ adds the diagram with exchanged external indices.}
\end{figure*}

In Fig.~\ref{fig:=CSigma(4)} we provide the diagrams for
$\Sigma$ to order $J^4$ with geometry factors and
$\sigma^{(nx)}$ given in Tabs.~\ref{tab:t4} and
\ref{tab:sigma_4}.

\begin{center}
\begin{table*}[h]
\begin{centering}
\begin{tabular}{c|l|l|l}
top. & r-space $t_{jj^\prime}^{(nx)}$ & k-space $t_{\mathbf{k}}^{(nx)}$ & nn-hyp. $\tilde{t}_{\mathbf{k}}^{(nx)}$\tabularnewline
\hline 
(4a) & $\delta_{jj^\prime}\sum_{i}J_{ij}^{4}$ & $\int_{\mathbf{p}_{1,2,3}}J_{\mathbf{p}_{1}}J_{\mathbf{p}_{2}}J_{\mathbf{p}_{3}}J_{\mathbf{p}_{1}+\mathbf{p}_{2}+\mathbf{p}_{3}}$ & $J_{1}^{3}\tilde{J}_{\mathbf{0}}$\tabularnewline
(4b) & $\delta_{jj^\prime}\left[\sum_{i}J_{ji}^{2}\right]^{2}$ & $\left[\int_{\mathbf{q}}J_{\mathbf{q}}^{2}\right]^{2}$ & $\left[J_{1}\tilde{J}_{\mathbf{0}}\right]^{2}$\tabularnewline
(4c) & $J_{jj^\prime}^{4}$ & $\int_{\mathbf{p}_{1,2,3}}J_{\mathbf{k}-\mathbf{p}_{1}}J_{\mathbf{p}_{2}}J_{\mathbf{p}_{3}}J_{\mathbf{p}_{1}+\mathbf{p}_{2}+\mathbf{p}_{3}}$ & $J_{1}^{3}\tilde{J}_{\mathbf{k}}$\tabularnewline
\hline 
(4d) & $\delta_{jj^\prime}\sum_{i}J_{ji}^{2}\left[J\cdot J\right]_{ij}$ & $\int_{\mathbf{q},\mathbf{p}}J_{\mathbf{q}}J_{\mathbf{p}}J_{\mathbf{q}+\mathbf{p}}^{2}$ & $0$\tabularnewline
(4e) & $J_{jj^\prime}^{2}\sum_{i}\left[J_{ji}^{2}+J_{j^\prime i}^{2}\right]$ & $2\left[\int_{\mathbf{p}}J_{\mathbf{p}}^{2}\right]\left[\int_{\mathbf{q}}J_{\mathbf{k}-\mathbf{q}}J_{\mathbf{q}}\right]$ & $2J_{1}^{2}\tilde{J}_{\mathbf{0}}\tilde{J}_{\mathbf{k}}$\tabularnewline
(4f) & $\delta_{jj^\prime}\sum_{i_{1,2}}J_{ji_{1}}^{2}J_{i_{1}i_{2}}^{2}$ & $\left[\int_{\mathbf{q}}J_{\mathbf{q}}^{2}\right]^{2}$ & $\left[J_{1}\tilde{J}_{\mathbf{0}}\right]^{2}$\tabularnewline
\hline 
(4g) & $J_{jj^\prime}^{2}\left[J\cdot J\right]_{jj^\prime}$ & $\int_{\mathbf{q},\mathbf{p}}J_{\mathbf{k}-\mathbf{p}}J_{\mathbf{p}-\mathbf{q}}J_{\mathbf{q}}^{2}$ & $0$\tabularnewline
(4h) & $\left[J^{2}\cdot J^{2}\right]_{jj^\prime}$ & $\left[\int_{\mathbf{q}}J_{\mathbf{k}-\mathbf{q}}J_{\mathbf{q}}\right]^{2}$ & $\left[J_{1}\tilde{J}_{\mathbf{k}}\right]^{2}$\tabularnewline
(4i) & $\delta_{jj^\prime}\left[J\cdot J^{2}\cdot J\right]_{jj}$ & $\int_{\mathbf{q}}\int_{\mathbf{p}}J_{\mathbf{q}}J_{\mathbf{p}}J_{\mathbf{q}+\mathbf{p}}^{2}$ & $0$\tabularnewline
\hline 
(4j) & $\delta_{jj^\prime}\left[J\cdot J\cdot J\cdot J\right]_{jj}$ & $\int_{\mathbf{q}}J_{\mathbf{q}}^{4}$ & $6d\left[2d-1\right]J_{1}^{4}$\tabularnewline
(4k) & $J_{jj^\prime}\left[J\cdot J^{2}+J^{2}\cdot J\right]_{jj^\prime}$ & $2\int_{\mathbf{q}}J_{\mathbf{k}-\mathbf{q}}J_{\mathbf{q}}\int_{\mathbf{p}}J_{\mathbf{q}-\mathbf{p}}J_{\mathbf{p}}$ & $0$\tabularnewline
(4l) & $\left[J\cdot J\right]_{jj^\prime}^{2}$ & $\int_{\mathbf{q}}J_{\mathbf{k}-\mathbf{q}}^{2}J_{\mathbf{q}}^{2}$ & $16J_{1}^{4}\left(\frac{d^{2}}{4}+\frac{1}{8}\sum_{\mu=1}^{d}\cos(2k_{\mu})+\sum_{\mu<\nu}^{d}\cos k_{\mu}\cos k_{\nu}\right)$\tabularnewline
\hline 
(4m) & $J_{jj^\prime}\left[J\cdot J\cdot J\right]_{jj^\prime}$ & $\int_{\mathbf{q}}J_{\mathbf{k}-\mathbf{q}}J_{\mathbf{q}}^{3}$ & $3\left[2d-1\right]J_{1}^{3}\tilde{J}_{\mathbf{k}}$\tabularnewline
\end{tabular}
\par\end{centering}
\caption{\label{tab:t4}Geometry factors $t^{(nx)}\left[J\right]$ for $\Sigma$-diagrams
of order $n=4$. See the caption of Tab.~\ref{tab:t023} for remarks.}
\end{table*}
\par\end{center}
\begin{table}
\begin{centering}
\begin{tabular}{c|l|l|l|l}
 & Ising $T^{1+n}\!\sigma^{zz(nx)}$ & TFIM $\sigma^{xx(nx)}(i\nu)|_{T=0}$ & Heisenberg: $T^{1+n}\sigma^{zz(nx)}(0)$ & Heisenberg: $T^{1+n}\sigma^{zz(nx)}(i\nu_{m}\neq0)$\tabularnewline
\hline 
(4a) & $\frac{1}{4!}b_{c,3}b_{c,5}$ & $\frac{-S^{2}\left(195h^{4}+38h^{2}\nu^{2}+3\nu^{4}\right)}{512h^{3}\left(h^{2}+\nu^{2}\right)^{2}\left(9h^{2}+\nu^{2}\right)}$ & $\frac{b_{c,1}^{2}}{-240}\left(\!192b_{c,1}^{3}\!+\!80b_{c,1}^{2}\!+\!20b_{c,1}\!+\!3\right)$ & $\frac{\Delta^{2}b_{c,1}^{2}}{15}\!\left(\!4\!\left[30\Delta^{2}\!+\!1\right]b_{c,1}\!+\!12b_{c,1}^{2}\!+\!15\Delta^{2}\!+\!2\!\right)$\tabularnewline
(4b) & $\frac{3}{4!}b_{c,1}^{2}b_{c,5}$ & $\frac{S^{3}\left(447h^{6}+405h^{4}\nu^{2}+93h^{2}\nu^{4}+7\nu^{6}\right)}{256h^{3}\left(h^{2}+\nu^{2}\right)^{3}\left(9h^{2}+\nu^{2}\right)}$ & $\frac{b_{c,1}^{3}}{24}\left(48b_{c,1}^{2}+12b_{c,1}+1\right)$ & $\frac{\Delta^{2}b_{c,1}^{3}}{-3}\left(6b_{c,1}+30\Delta^{2}+1\right)$\tabularnewline
(4c) & $0$ & $0$ & $\frac{b_{c,1}^{2}}{-120}\left(12b_{c,1}^{2}+6b_{c,1}+1\right)$ & $\frac{\Delta^{2}b_{c,1}^{2}}{-15}\!\left(\!4\!\left[30\Delta^{2}\!+\!1\right]b_{c,1}\!+\!12b_{c,1}^{2}\!+\!15\Delta^{2}\!+\!2\!\right)$\tabularnewline
\hline 
(4d) & $0$ & $0$ & $\frac{b_{c,1}^{3}}{24}\left(4b_{c,1}+1\right)$ & $\frac{\Delta^{2}b_{c,1}^{3}}{-2}$\tabularnewline
(4e) & $0$ & $0$ & $\frac{b_{c,1}^{3}}{48}\left(4b_{c,1}+1\right)$ & $\frac{\Delta^{2}b_{c,1}^{3}}{3}\left(6b_{c,1}+30\Delta^{2}+1\right)$\tabularnewline
(4f) & $\frac{6}{4!}b_{c,1}b_{c,3}^{2}$ & $\frac{S^{3}\left(21h^{2}+5\nu^{2}\right)}{256h^{3}\left(h^{2}+\nu^{2}\right)^{2}}$ & $b_{c,1}^{3}\left(b_{c,1}^{2}+\frac{b_{c,1}}{3}+\frac{1}{30}\right)$ & $\frac{\Delta^{2}b_{c,1}^{3}}{-3}\left(6b_{c,1}+12\Delta^{2}+1\right)$\tabularnewline
\hline 
(4g) & $\frac{12}{4!}b_{c,1}b_{c,3}^{2}$ & $\frac{S^{3}\left(21h^{2}+\nu^{2}\right)}{16h\left(h^{2}+\nu^{2}\right)^{2}\left(9h^{2}+\nu^{2}\right)}$ & $\frac{b_{c,1}^{3}}{120}\left(144b_{c,1}^{2}+48b_{c,1}+5\right)$ & $\frac{b_{c,1}^{3}}{3}\Delta^{2}\left(12\Delta^{2}+1\right)$\tabularnewline
(4h) & $0$ & $0$ & $\frac{b_{c,1}^{3}}{120}$ & $-6\Delta^{2}b_{c,1}^{3}$\tabularnewline
(4i) & $0$ & $0$ & $\frac{b_{c,1}^{3}}{120}\left(10b_{c,1}+1\right)$ & $\frac{b_{c,1}^{3}}{6}\Delta^{2}\left(24\Delta^{2}-1\right)$\tabularnewline
\hline 
(4j) & $\frac{12}{4!}b_{c,1}^{3}b_{c,3}$ & $\frac{-5S^{4}\left(11h^{2}+3\nu^{2}\right)}{256h^{3}\left(h^{2}+\nu^{2}\right)^{2}}$ & $\frac{b_{c,1}^{4}}{-6}\left(6b_{c,1}+1\right)$ & $+2\Delta^{2}b_{c,1}^{4}$\tabularnewline
(4k) & $0$ & $0$ & $\frac{b_{c,1}^{3}}{-120}$ & $\frac{b_{c,1}^{3}}{-6}\Delta^{2}\left(24\Delta^{2}-1\right)$\tabularnewline
(4l) & $0$ & $0$ & $0$ & $0$\tabularnewline
\hline 
(4m) & $0$ & $0$ & $0$ & $-2\Delta^{2}b_{c,1}^{4}$\tabularnewline
\end{tabular}
\par\end{centering}
\caption{\label{tab:sigma_4}Ising model, TFIM at $T=0$ and Heisenberg model:
The lattice independent part $\sigma^{(nx)}$ for $n=4$ for all topologies, c.f.~Eq.~(\ref{eq:Sigma=t*sigma}).
The Ising case is purely static.}
\end{table}
\FloatBarrier
\twocolumngrid

\bibliography{MyLibrary}

\end{document}